\documentclass{article}

\usepackage{arxiv}

\usepackage[utf8]{inputenc} 
\usepackage[T1]{fontenc}    
\usepackage{hyperref}       
\usepackage{url}            
\usepackage{booktabs}       
\usepackage{amsfonts}       
\usepackage{nicefrac}       
\usepackage{microtype}      
\usepackage{lipsum}		
\usepackage{graphicx}
\usepackage{natbib}
\usepackage{doi}

\usepackage{amsmath}
\usepackage{threeparttable}
\usepackage{multirow}
\usepackage{graphicx}
\usepackage{subfig}
\usepackage{float}
\graphicspath{ {./figures/} }
\usepackage{adjustbox,xr,comment} 

\title{Getting it right: Methods for risk ratios and risk differences  cluster randomized trials with a small number of clusters}

\date{November 20, 2025}	

\author{
  Shifeng Sun \\
  Department of Biostatistics and Bioinformatics\\
  Duke University\\
  Duke Global Health Institute\\
  Durham, North Carolina, USA\\
  \And
  Xueqi Wang \\
  Department of Biostatistics\\
  Yale School of Public Health, Yale University\\
  New Haven, Connecticut, USA\\
  \And
  Zhuoran Hou \\
  Department of Biostatistics and Bioinformatics\\
  Duke University\\
  Durham, North Carolina, USA\\
  \And
  Elizabeth L.~Turner \\
  Department of Biostatistics and Bioinformatics\\
  Duke University\\
  Duke Global Health Institute\\
  Durham, North Carolina, USA\\
  \texttt{liz.turner@duke.edu}
}

\renewcommand{\shorttitle}{\textit{arXiv} Template}

\hypersetup{
pdftitle={A template for the arxiv style},
pdfsubject={q-bio.NC, q-bio.QM},
pdfauthor={David S.~Hippocampus, Elias D.~Striatum},
pdfkeywords={First keyword, Second keyword, More},
}

\begin{document}
\maketitle

\begin{abstract}
Most cluster randomized trials (CRTs) randomize fewer than 30-40 clusters in total. When performing inference for such ``small'' CRTs, it is important to use methods that appropriately account for the small sample size. When the generalized estimating equations (GEE) approach is used for analysis of ``small'' CRTs, the robust variance estimator from GEE is biased downward and therefore bias-corrected standard errors should be used. Moreover, in order to avoid inflated Type I error, an appropriate bias-corrected standard error should be paired with the t- rather than Z-statistic when making inference about a single-parameter intervention effect. Although several bias-correction methods (including Kauermann and Carroll (KC), Mancl and DeRouen (MD), Morel, Bokossa, and Neerchal (MBN), and the average of KC and MD (AVG)) have been evaluated for inference for odds ratios, their finite-sample behavior in ``small'' CRTs with few clusters has not been thoroughly investigated for risk ratios and risk differences. The current article aims to fill the gap by including analysis via binomial, Poisson and Gaussian models and for a broad spectrum of scenarios. Analysis is via binomial and Poisson models (using log and identity link for risk and differences measures, respectively). We additionally explore the use of Gaussian models with identity link for risk differences and adopt the "modified" approach for analysis with misspecified Poisson and Gaussian models. We consider a broad spectrum of scenarios including  for rare outcomes, small cluster sizes, high intracluster correlations (ICCs), and high coefficients of variation (CVs) of cluster size. 
\end{abstract}


\section{Introduction}
Cluster randomized trials (CRTs), also known as group randomized trials, are studies that randomize groups or clusters of individuals (e.g. schools, hospitals or communities) and measure individual-level outcomes. Although the effective sample size of CRTs is lower than a standard individually-randomized trial with the same overall sample size due to the expected correlation of outcomes of individuals in the same cluster, they are the design of choice when the intervention is at the cluster level  \cite{murray2020essential} such as a community-wide physical activity program \cite{kamada2018community} or a school-directed nutritional and physical activity policy for school-aged children \cite{ickovics2019implementing}. Similarly, the CRT is the design of choice when there is the potential for intervention contamination. For example, this could arise in the case that individual-level randomization were to be used to evaluate a new psychological therapy delivered by a therapist and the same therapist delivered both the new therapy and the comparison therapy. In this case, cluster randomization of therapists (with all their patients allocated to the same trial arm) would minimize the spread of the new therapy to individuals not working with a therapist randomized to deliver the new therapy \cite{murray2020essential}.

Binary outcomes are are often selected to measure intervention effectiveness in CRTs \cite{fiero2016statistical}. Such outcomes are typically analyzed using regression-based methods that account for correlation of individual-level outcomes. The most commonly-used methods are  generalized linear mixed models (GLMM) and generalized estimating equations (GEE) \cite{turner2021completeness}. Whereas the GLMM approach directly models the assumed outcome correlation structure through the use of random effects, the GEE approach separately estimates the mean and variance structure. Benefits of the GEE approach include the population-averaged interpretation of the estimated intervention effects, rather than the cluster-specific interpretation of effects estimated via GLMM, as well as robustness to misspecification of the correlation structure in large samples \cite{preisser2003integrated}. For this reason, in the current paper, we focus on the GEE approach for the analysis of binary outcomes in CRTs.

Despite the benefits of the GEE approach, just like the GLMM approach, the GEE approach faces challenges in performing valid inference for  “small” CRTs, namely those with fewer than 40 clusters. More specifically,  because of finite-sample bias,  the robust sandwich variance estimator is expected to provide considerably inflated Type I error rates for inference regarding the intervention effect in small CRTs \cite{feng2001selected} \cite{mancl2001covariance} \cite{leyrat2018cluster}. Such a limitation is important given that a recent systematic review of CRTs with primary binary outcome demonstrated that most {(71\%)} CRTs had fewer than 40 clusters \cite{turner2021completeness}.

To address the potential for invalid inference from small CRTs analyzed via GEE, multiple correction methods to robust SEs have been proposed, including by Kauermann and Carrol (KC) \cite{kauermann2001note}
; Mancl and DeRouen, (MD) \cite{mancl2001covariance}
; Fay and Graubard (FG) \cite{fay2001small} Morel et al., (MBN) \cite{morel2003small}
and by Ford and Westgate (AVG; the average of {the Kauermann and Carrol and Mancl and DeRouen})) \cite{ford2020maintaining}. Using simulation studies, several authors have   evaluated the performance of these small-sample correction methods for the case of analysis of a parallel two-arm CRT with binomial outcome distribution and logit link, that is, for a standard logistic-regression GEE analysis of correlated binary outcomes used to estimate odds ratios (ORs).\cite{li2015small} \cite{westgate2013small} \cite{wang2016covariance} Li and Redden \cite{li2015small}  evaluated the Type I error performance of inference based on both a Wald $z$-test and a Wald $t$-test  each combined with four different bias-corrected SEs, and where degrees of freedom for the $t$-test are given by $N - p$, for $N$ the total number of clusters and $p$ the number of cluster-level parameters in the mean model of the GEE analysis (typically two, namely an intercept and intervention-arm indicator). This was  at an outcome proportion of 0.25 and with balanced and variable cluster size. In the scenarios considered (Table 1 and 2), the simulation study showed that inference based on the Wald t-test was preferred over the z-test based on Type I error performance and, more specifically, KC-corrected SEs were preferred  for designs with balanced (i.e. equal) cluster sizes or those with small variation in cluster size, while FG-corrected SEs were preferred when there was larger variation in cluster size. 

In contrast to the evaluation of small-sample corrections methods for the estimation of odds ratios (OR) \cite{li2015small},  the evaluation of methods for  risk ratios (RR) and risk differences (RD) has been more limited in scope with no consideration of the FG or MBN approaches \cite{yelland2011performance} \cite{zou2004modified} \cite{pedroza2016performance} \cite{pedroza2017estimating}.  Given that the CONSORT statement on the reporting of trials recommends that the intervention effect be reported using both absolute measures (eg, RD)  and relative measures (eg, RR and OR) of effect \cite{schulz1996academia} \cite{campbell2004consort}, there is a need to address this gap in the literature. 
 
The current manuscript aims to fill this gap by evaluating the model performance (i.e. convergence and percent bias) of a range of analytical models to estimate risk ratios (RR) and risk differences (RD) using GEE analysis of binary outcome data from cluster randomized trials and to evaluate the Type I error performance of the Wald t-test for the intervention effect (RR and RD)  for five small-sample correction methods applied to the sandwich variance estimator. More specifically, the models are: (1) two models for RRs: log-binomial GEE and log-modified Poisson GEE, and, (2) three models for RDs:  identity-binomial GEE, identity-modified Poisson GEE and identity-modified Gaussian GEE. Here we use the term "modified" (e.g. modified Poisson GEE) to refer to the case where the outcome distribution is mis-specified (i.e. Poisson is misspecified for binary outcomes) but the SEs are "modified" such that correction for the outcome misspecification is achieved via robust sandwich variance estimators (see details in methods below). We note that the term "modified Poisson" has been previously used in the literature to refer specifically to the case with a log link. Here we use it more generally to refer to the case where the outcome distribution is misspecified (and for which robust SEs are used to correct for that misspecification) for any choice of link function. As such, we use "log-modified Poisson" to distinguish from the identity link, i.e. "identity-modified Poisson". Similarly, we therefore use "identity-modified Gaussian"  to refer to a model with identity link where the outcome distribution is misspecified as Gaussian and robust SEs are used. 

The manuscript is organized as follows: Section 2 describes the GEE methodology for the analysis of binary outcomes under a range of assumed outcome distributions and link functions, together with the form of the five small-sample correction methods considered. Section 3 outlines a simulation study together with findings of that simulation study to evaluate performance of the selected models and the Type I error performance of the t-test paired with each small-sample correction methods. Section 4 presents a comparative analysis of a small CRT to show the potential impact of ignoring the small-sample correction methods and how findings may change when a valid small-sample correction is used. Section 5 ends with a discussion.

\begin{table}[htbp]
    \begin{adjustbox}{max width=\textwidth}
    \begin{threeparttable}
    \caption{Studies including small sample corrections for parallel two-arm cluster trials}
    \label{tab:my_label}
    \centering
    \begin{tabular}{|c|c|c|c|c|c|c|c|c|c|c|c|c|c}
    \hline
        \multirow{2}{*}{\begin{tabular}{c}Reference\end{tabular}}
        & \multicolumn{3}{c|}{Measure\tnote{a}} & \multicolumn{5}{c|}{Small Sample Correction\tnote{b}} & \multicolumn{2}{c|}{Performance Measure} \\
        \cline{2-11} & OR & RR & RD  & KC & MD & FG & MBN & AVG & Coverage & Type I error \\
    \hline
        Westgate (2013) & X & - & - & X & X & - & - & - & - & X \\
        Li (2015) & X & - & -       & X & X & X & X & - & - & X \\
        Wang (2015) & X & - & -     & X & X & X & X & - & - & X \\
        Ford (2017)  & X & - & -      & X & X & X & - & X & - & X \\
        Ford (2017)  & X & - & -      & X & X & X & - & X & - & X \\
         Zhu (2023) & X & - & -     & - & X & - & - & - & X & - \\
        Yelland (2011) & - & X & -  & - & X & - & - & - & X & X \\
        Zou (2013) & - & X & -      & - & X & - & - & - & X & - \\
        Li (2021)  & - & X & -      & X & X & X & - & X & X & - \\
        Pedroza (2017) & - & X & -  & X & - & - & - & - & X & - \\
        Pedroza (2016) & - & - & X  & - & X & - & - & - & X & - \\
        Jules (2024) & - & - & X  & - & - & X & - & - & X & X \\
    \hline
    \end{tabular}
            \begin{tablenotes}
               \item [a] RR: Risk Ratio; OR: Odds Ratio; RD: Risk Difference.
               \item [b] KC: Kauermann and Carrol;MD: Mancl and DeRouen, MD;FG, Fay and Graubard; MBN: Morel, Bokossa and Neerchal; AVG, average of KC and MD.
            \end{tablenotes}
    \end{threeparttable}
    \end{adjustbox}
\end{table}

\begin{table}[htbp]
    \begin{adjustbox}{max width=\textwidth}
    \begin{threeparttable}
    \caption{Simulation factors in studies of Table1}
    \label{tab:my_label}
        \begin{tabular}{|l|c|c|c|c|c|c|}
            \hline
            Reference & Total clusters & Cluster size & Variation\tnote{a} & Outcome proportion & ICC\tnote{b} & Covariate\tnote{c} \\
            \hline
            OR & & & & & &\\
            \hspace{3mm} Westgate (2013) & 12, 20, 40 & 25 & X & 0.05, 0.10, 0.20, 0.50 & 0.05, 0.10 & X \\
            \hspace{3mm} Li (2015) & 10, 20, 30 & 50, 100, 150 & X & 0.25 & 0.001, 0.01, 0.05 & - \\
            \hspace{3mm} Wang (2015) & 10,20,30,40,50 & 5, 10, 20 & - & N/A & 0.1 & - \\
            RR & & & & & &\\
            \hspace{3mm} Ford (2017) & 10,20 & 50, 100 & X & 0.25 & 0.01, 0.05, 0.1 & X \\
            RR & & & & & &\\
            \hspace{3mm} Yelland (2011) & 20, 50 & 25, 10 & X & 0.1, 0.2 & 0.01, 0.15 & X \\
            \hspace{3mm} Zou (2013) & 50, 100 & 7.5 & X & 0.2, 0.4 & 0.01, 0.05, 0.30 & X \\
            \hspace{3mm} Pedroza (2017) & 4, 10, 30 & 10, 20, 50 & - & 0.3 & 0.08 & X \\
            \hspace{3mm} Li (2021) & 11, 21, 33, 46, 59 & 50, 100 & X & 0.15, 0.3 & 0.01, 0.05, 0.1, 0.15, 0.2 & - \\
            RD & & & & & &\\
            \hspace{3mm} Pedroza (2016) & 18 & 10, 50, 100 & - & 0.1, 0.25, 0.5 & 0.01, 0.05, 0.1 & X \\
            \hspace{3mm} Jules (2024) & 10, 20, 40, 60, 80 & 10 to 280\tnote{d} & X & 0.1, 0.2, 0.5 & 0.001, 0.01, 0.05, 0.2 & X \\
            \hline
        \end{tabular}
    \begin{tablenotes}
    \item [a] Variation: Cluster size variation.
    \item [b] ICC: Intra-cluster correlation coefficient.
    \item [c] Covariate: Baseline covariate included in the data generation mechanism. 
    \item [d] Depending on total number of clusters
    \end{tablenotes}        
    \end{threeparttable}
    \end{adjustbox}
\end{table}

\section{Methods}

Consider a parallel two-arm cluster randomized trial consisting of $N$ clusters with cluster size of $M_i$, for $i=1,...,M_i$. Let $Y_{ij}$ be the binary outcome for the $j$th individual, $j=1,...,M_i$, in the $i$th cluster  and {$\mathbf{X_{ij}}$}  a vector of individual-level and cluster-level covariates. For simplicity, let {$\mathbf{X_{ij}}$} include only an intercept and a cluster-level intervention indicator (equal to 1 if the cluster is assigned to the intervention arm or 0 if the cluster is assigned to the control arm) such that $\mathbf{X_{ij}} = (1,0)'$ for individuals in control-arm clusters and $\mathbf{X_{ij}} = (1,1)'$ for individuals in intervention-arm clusters. Let $\mu_{ij}=E(Y_{ij}|\mathbf{X_{ij}})$ be the marginal mean outcome given $\mathbf{X_{ij}}$. We note that this is assumed to be the mean of a binomial distribution using standard logistic-binomial regression or the mean of a Poisson distrbution or Gaussian distribution, under the modified-Poisson and modified-Gaussian approaches, respectively.  The GEE model specifying their association can be written by the following generalized linear model:     
\[g(\mu_{ij})=\mathbf{X_{ij}\beta}\]
where $g$ is a link function and $\mathbf{\beta}$ is a $2\times1$ vector of regression parameters. The link functions considered in the current article are the canonical: the log and the identity link, such that $g$ can be written as:
\begin{align*}
g(\mu_{ij})&=log(\mu_{ij})\\
g(\mu_{ij})&=\mu_{ij} 
\end{align*}
respectively. 

The regression parameter estimator $\mathbf{\hat{\beta}}$ solves the following generalized estimating equations
\[\sum_{i=1}^{N}\mathbf{D_i^{'}V_i^{-1}(Y_i-\mu_i)=0}\]
where $\mathbf{D_i=\frac{\partial \mu_i}{\partial \beta^{'}}}$ and $\mathbf{V_i}$ is the working covariance matrix for $\mathbf{Y_i}$. More specifically, $\mathbf{V_i=A_i^{1/2}R_i(\alpha)A_i^{1/2}}$
with $A_i=Diag\{V(\mu_{i1}),...,V(\mu_{in_i})\}$ and the working correlation matrix $\mathbf{R_i(\alpha)}$ describes the correlation pattern of observations within the same cluster. In this study, we specify $\mathbf{R_i(\alpha)}$ as exchangeable, that is $\alpha = \rho$ for all pairs of observations in the same clusters, with outcomes of individuals in different clusters uncorrelated. More specifically, $Corr(Y_{ij}, Y_{i'k})=\begin{cases} \rho & {i=i', j\neq k} \\ 0 & {i \ne i'}  \end{cases}$. The robust variance-covariance matrix of $\mathbf{\hat{\beta}}$ is estimated by 
\[ \mathbf{V_{robust}}=\mathbf{\hat{\sum}_1^{-1}}\mathbf{\hat{\sum}_0^{-1}}\mathbf{\hat{\sum}_1^{-1}}\]
where 
\[\mathbf{\hat{\sum}_0=N^{-1}\sum_{i=1}^N C_i D_i^{'}(\hat{\beta}) V_i^{-1}(\hat{\alpha}) (Y_i-\hat{\mu}_i) (Y_i-\hat{\mu}_i) V_i^{-1}(\hat{\alpha}) D_i(\hat{\beta}) C_i^{'}} \] 
and \[\mathbf{\hat{\sum}_1^{-1}}=\mathbf{V_{model}}=\{N^{-1}\sum_{i=1}^N D_i^{'}(\hat{\beta}) V_i^{-1}(\hat{\alpha}) D_i(\hat{\beta})\}^{-1}.\]
To address potential negative bias of the robust variance estimator in the small sample setting, several methods propose adjusting the multiplicative factor $\mathbf{C_i}$. First, define $Q_i=D_i^{'}V_i^{'}D_i(N\sum_1)^{-1}$. Setting $C_i=(I_p-Q_i)^{-1/2}$ gives the Kauermann-Carrol (KC) variance estimator \cite{kauermann2001note}. Setting $C_i=(I_p-Q_i)^{-1}$ gives the Mancl-DeRouen (MD) variance estimator \cite{mancl2001covariance}. Setting $C_i=\text{diag}\left[\{1-\min(r, [Q_i]_{jj})\}^{-1/2}\right]$ where $r=0.75$ by default and can also be user-defined with an upper bound of 1 gives the Fay-Graubard (FG) variance estimator \cite{fay2001small}. The AVG variance estimator is calculated as an average of KC and MD variance estimators, which has been shown to perform better in terms of bias and type I error rate \cite{ford2020maintaining}. In addition to multiplicative adjustments, there is also an additive adjustment due to the Morel-Bokossa-Neerchal (MBN) variance estimator, given by $\mathbf{V_{MBN}}=c\mathbf{V_{robust}}+\mathbf{\delta_N \phi \sum_1^{-1}}$, where $c=\left\{\left(\sum_{i=1}^{N}N_i-1\right)/\left(\sum_{i=1}^{N}N_i-2\right)\right\}\times\left\{N/(N-1)\right\}$, $\delta_N = \min\{0.5, 2/(N-2)\}$, and $ \phi = \text{max}\left[1, \frac{\text{trace}\left(\mathbf{c}\left(\sum_{i=1}^N D_i^{'} V_i^{-1}(\hat{\alpha}) (Y_i-\hat{\mu}_i) (Y_i-\hat{\mu}_i) V_i^{-1}(\hat{\alpha}) D_i\right) (N\sum_1)^{-1}\right)}{2}\right] $ \cite{morel2003small}.

\section{Simulation Study} 
To evaluate the performance of the five bias-corrected robust SEs described above for GEE inference of risk differences and risk ratios in parallel two-arm CRTs, we undertook a simulation study. Here we outline the design of our factorial simulation study using the ADEMP framework \cite{morris2019using}.  
\subsection{Study Description}
\textbf{Aims}: Our primary goals are to assess the Type I error performance of a range of published small-sample correction methods for robust standard errors (SEs) of two different intervention effect measures (RR and RD) for a parallel two-arm cluster randomized trial with binary outcome and to assess convergence properties of the different analytical methods used. For comparative purposes, we also considered inference for the OR effect measure (with all results presented in supplementary materials) in order to expand the earlier study to a broader range of outcome risk levels. In particular, we considered one, two and three different analytic models for the OR, RR and RD, respectively, selected based on previously published methods. We note that, as described in the background, although some of the proposed methods have been evaluated for each of the three measures, no single simulation study has addressed all of these models and small-sample correction methods. While all the models above are assessed for balanced design, the unbalanced design is also considered for modified Poisson model with either log or identity link to test its robustness. 

\textbf{Data-generating mechanism}: Correlated binary outcome data were generated separately for each cluster using the method of Qaqish \cite{qaqish2003family}, based on an assumed marginal mean and the exchangeable working correlation matrix. More specifically, the following marginal mean model was assumed for each observation $j$ in cluster $i$. In order to mimic the setting with a small number of clusters, and to build from the parameters assumed in the earlier simulation study of Li and Redden \cite{li2015small}, the total number of clusters (N) was chosen from 10, 20, 30, 40 and 50, with non-variable cluster size (M) for a given scenario of 10, 30, 50 and 100, where it was assumed that $N/2$ clusters were allocated to each arm (i.e. balanced allocation). The outcome proportion ($\pi_0$) was set to be 0.02, 0.05, 0.1, 0.3 and 0.5 to cover different prevalences commonly encountered in public health, and the outcome proportions were assumed to be equal in the two arms ($\pi_1 = \pi_0$) in order to evaluate the empirical Type I error rate. (We note that these outcome proportions extend those considered in the earlier study by Li and Redden \cite{li2015small} to the rare outcome setting i.e. $\pi_0= 0.05$). The intra-cluster correlation coefficient (ICC) was selected from $\rho=0.01, 0.05$ and $0.1$, representing different levels of pairwise within-cluster correlation, where we note that we used the ICC for a binary outcome on the natural proportions scale (i.e. that which is required in order to implement the data generating method of Qaqish et al. \cite{qaqish2003family}). A total of 300 simulation scenarios (5 total number of clusters $\times$ 4 cluster sizes $\times$ 5 outcome proportions $\times$ 3 ICCs) were investigated under the framework of a full factorial simulation design i.e. in a $5\times4\times5\times3$ design. For unbalanced design, the cluster sizes were generated from a gamma distribution with mean equal to $\Bar{M} \in \{30, 100\}$ and CV ranging from $\{0.25, 0.50, 0.75, 1\}$ for each scenario (the non-integer values were rounded to the nearest integer), with the outcome proportion in \{0.05, 0.3 \} and ICC remaining the same. 

\textbf{Estimand}: Type I error performance, quantified based on the empirical Type I error proportion, model convergence and relative bias in SE estimation.  

\textbf{Methods}: Each scenario was analyzed using six different GEE models in order to estimate three different intervention effect estimates based on different combinations of the assumed outcome distribution and the assumed link, each with mean model specified with intercept and intervention-arm indicator only. Specifically: (1) for RR, two approaches were used, namely the log-binomial GEE and log-modified Poisson GEE; (2) for RD, three approaches were used, namely the identity-binomial GEE, modified-identity Poisson  GEE; and modified-identity Gaussian GEE. Given that only the logit link combined with the binomial distribution is canonical, lack of convergence is expected for all other non-canonical approaches and therefore we additionally provide estimates of convergence. For each of six models, seven standard error estimators were considered: model-based (MB), robust (i.e. based on the standard sandwich variance) and the following five corrections to robust SEs (KC, MD, FG, MBN and AVG).

\textbf{Performance measures}: For each scenario, the empirical type I error rate was calculated using 1000 simulated datasets, and the convergence rate was also calculated. More specifically, the empirical type I error rate was calculated based on the Wald t-test with $N-2$ degrees of freedom, and then compared with the nominal type I error rate of 5\%. An empirical type I error rate from 3.6\% to 6.4\% was considered acceptable based on 1000 replicates. The percent bias of empirical standard errors using different variance estimators was calculated by $(1/1000)\sum_{i=1}^{1000}(\hat{SE_i}-ESD)/ESD\times100$, where ESD denotes the empirical standard deviation of the $\hat{\beta_1}$ values resulting from the 1000 simulations in a given setting. All simulations and analyses were performed in Rstudio 4.1.1. \emph{geem} function from the \emph{geeM} package was used to fit all GEE models. All the variance estimators were estimated using R code adapted from a four-level CRT study \cite{wang2022power}. 

\subsection{Study Results}

\subsubsection{Convergence rate}
Convergence rates for all models with total number of clusters (n) $\geq20$ and non-rare outcomes ($\pi_0 \geq 0.1$) exceeded 0.9 (see Supplemental Material). The convergence rates of all models were adversely affected by rare outcomes, small cluster sizes, and a small number of total clusters. In contrast, the intraclass correlation coefficient (ICC) had minimal influence on convergence rates except when the total number of clusters was 10 or 20 with rare outcome. Regardless of whether the log link or the identity link was used, the Poisson model demonstrated convergence rates that were not lower than those of the binomial model in most scenarios (227/300 for both the log link and the identity link), highlighting its relative advantage. For the identity link under rare outcomes ($\pi_0 \leq 0.05$), both the Poisson and binomial models had convergence lower than 0.6 whereas convergence under the Gaussian model exceeded 0.67 in these instances.

\subsubsection{Relative Bias}
The robust estimators exhibited a downward bias in small samples across most scenarios (Fig1-2, Supp FigS1(a) and FigS2(a). All other estimators included in this study could partially correct this bias to varying degrees. Among them, the KC estimator typically resulted in the smallest absolute bias values, while others tended to overcorrect the bias. However, when the outcome proportion was 0.02 and the cluster size was 10, the robust estimator showed an upward bias, leading to even more biased results from the corrective estimators. This observation is consistent with findings in the Type I error rates.

\subsubsection{Type I error rate}
When employing the Poisson-log model in a balanced design, the robust estimator resulted in inflated Type I error rates when there were fewer than 30 clusters in most scenarios (Figure 3, Supp FigS1(b). However, it surprisingly maintained nominal Type I error rates when the outcome was rare ($\pi_0$ $\leq$ 0.1) and the cluster size was small (m $\leq$ 10). In situations where the robust estimator resulted in inflated Type I error rates, the KC estimator emerged as a suitable alternative due to its appropriate test size, compared to the conservative test sizes of other estimators. Nevertheless, in cases of higher ICC (ICC $\geq$ 0.05), nearly all estimators, including the KC, exhibited dramatically inflated Type I error rates. Therefore, more conservative estimators, such as the MD, FG, and MBN, are recommended. Similar findings were observed when using the binomial-log model, with Type I error rates showing greater instability in small sample settings as cluster sizes varied (Supp FigS3(a-c)). Given the unique impact of rare outcomes and the common use of the odds ratio for binary outcomes, the binomial-logit model was also fitted, yielding consistent conclusions with the first two models (Supp FigS6(a-c)). 

Many of the findings from the Poisson-log model were applicable to the Gaussian-identity model as well (Figure4, Supp FigS2(a-b)). However, in the presence of higher ICC (ICC $\geq$ 0.05), the robust estimator did not inflate Type I error rates when the outcome was rare ($\pi_0$ $\leq$ 0.1), regardless of cluster size. In scenarios where the robust estimators failed to maintain nominal test sizes, the KC estimator consistently served as a proper substitute. This conclusion also extends to the binomial-identity and Poisson-identity models (Supp FigS4-S5(a-c)). 

When the cluster size varied, the Type I error rate was also inflated for rare outcomes with a small average cluster size (m = 30) in the Poisson-log model. The KC estimator could be used as an alternative unless the outcome was rare combined with a large average cluster size or a high coefficient of variation (CV = 1.0). In such cases, the MD estimator should be employed. Regarding the Gaussian-identity model, the robust estimator maintained the Type I error rate when the outcome was rare with a large cluster size. For other scenarios under Gaussian-identity model, the KC estimators remained ideal.

\begin{figure}[htp]
    \centering
    \subfloat[Poisson model with log link, ICC=0.01]{%
        \includegraphics[clip, width=0.9\textwidth]{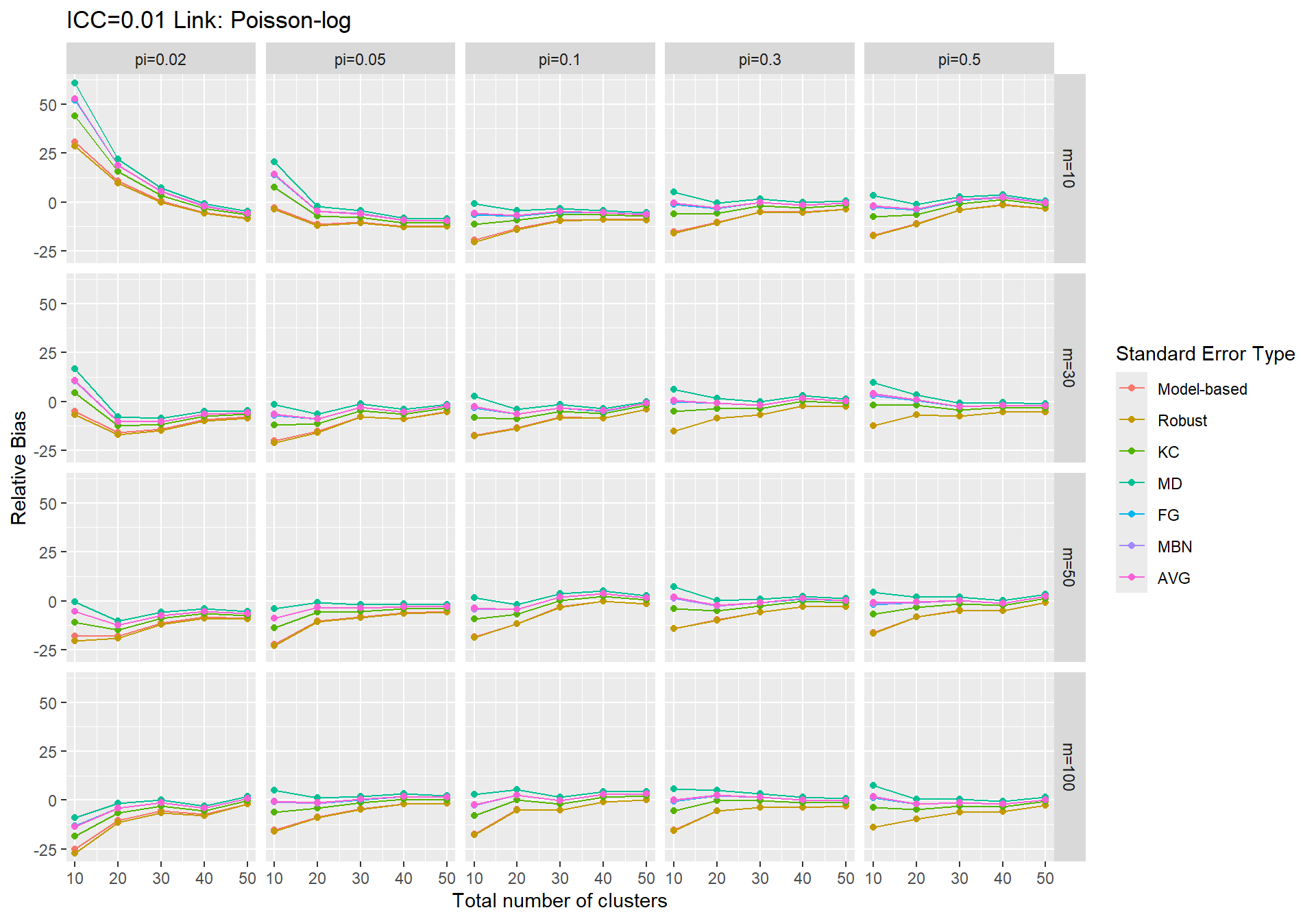}
        } 
    
    \subfloat[Poisson model with log link, ICC=0.1]{%
        \includegraphics[clip, width=0.9\textwidth]{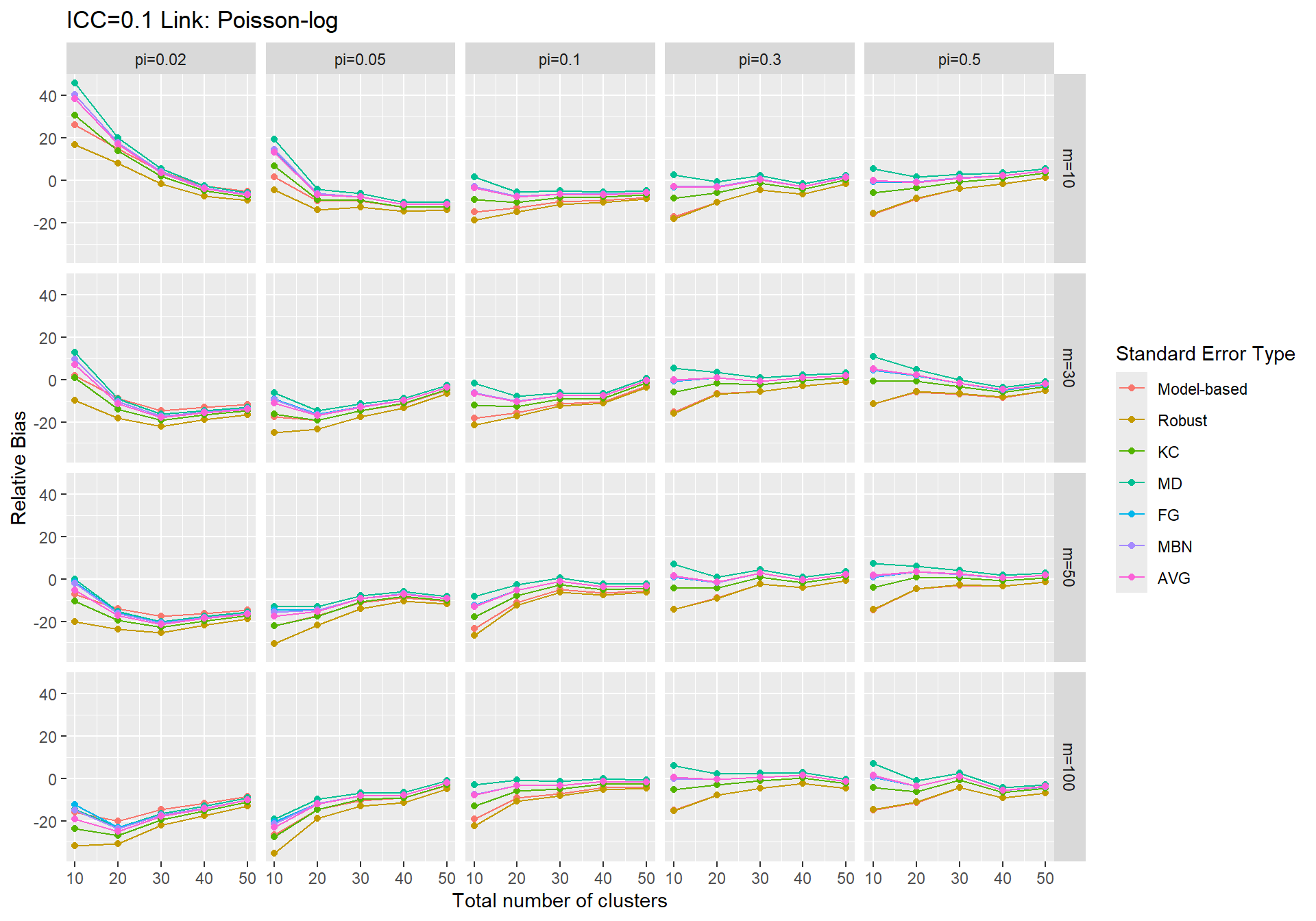}
        }
    \caption{Relative bias of GEE analyses using Poisson model to estimate risk ratios in balanced clustered data}
    \label{fig:my_label}
    \centering
\end{figure}

\begin{figure}[htp]
    \centering
    \subfloat[Gaussian model with identity link, ICC=0.01]{%
        \includegraphics[clip, width=0.9\textwidth]{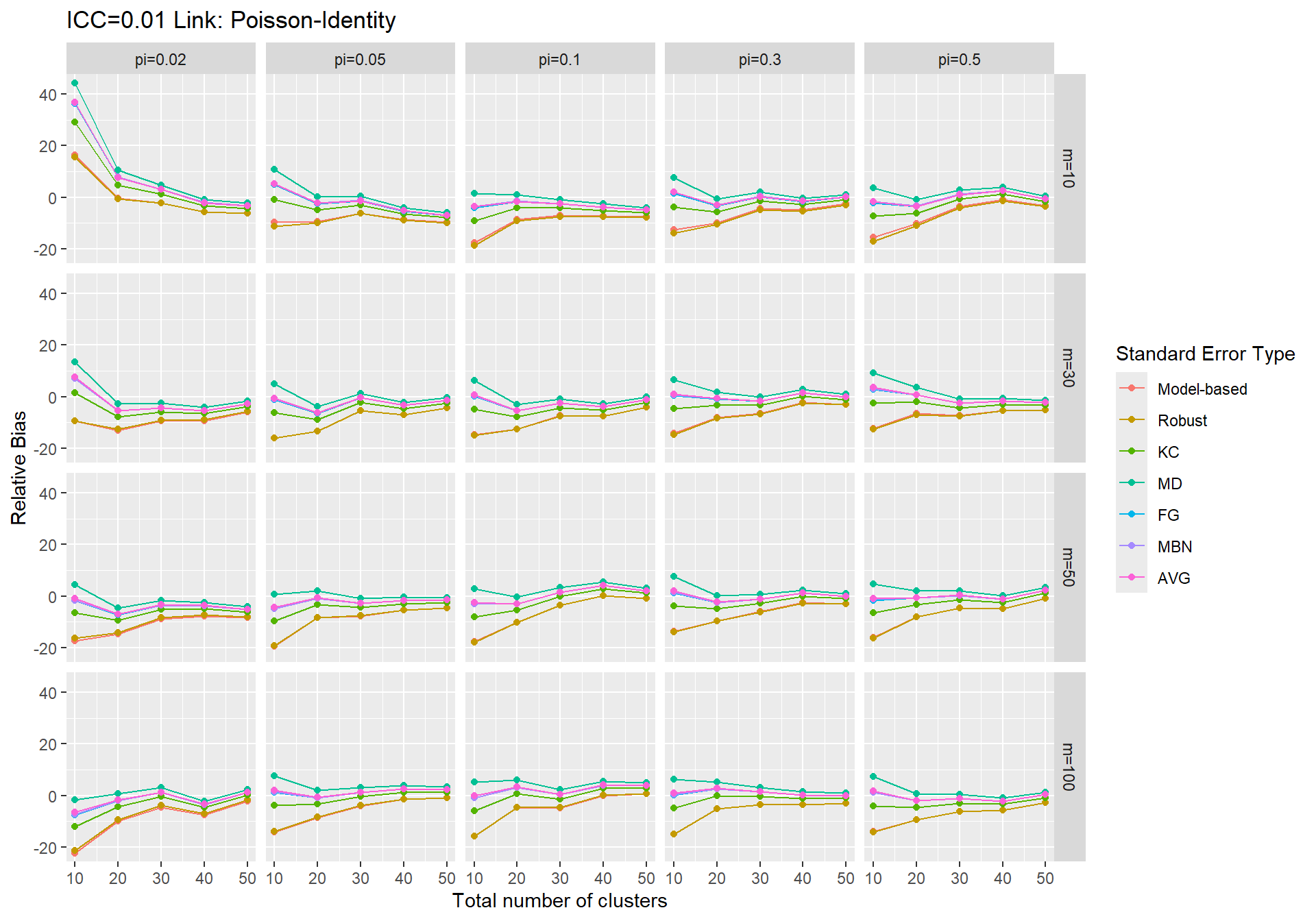}
        } 
    
    \subfloat[Gaussian model with identity link, ICC=0.1]{%
        \includegraphics[clip, width=0.9\textwidth]{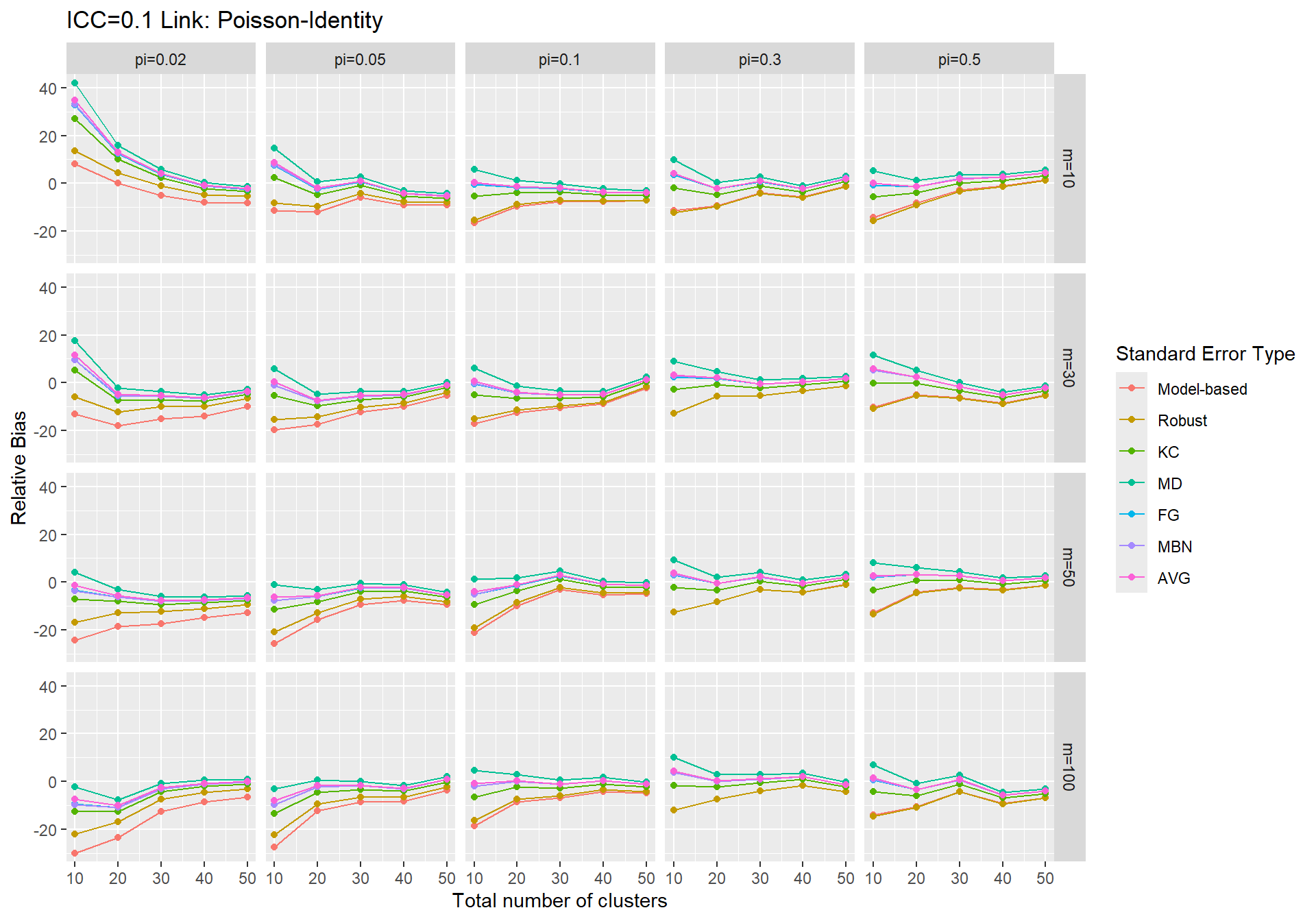}
        }
    \caption{Relative bias of GEE analyses using Poisson model to estimate risk differences in balanced clustered data}
    \label{fig:my_label}
    \centering
\end{figure}

\begin{figure}[htp]
    \centering
    \subfloat[Poisson model with log link, ICC=0.01]{%
        \includegraphics[clip, width=0.9\textwidth]{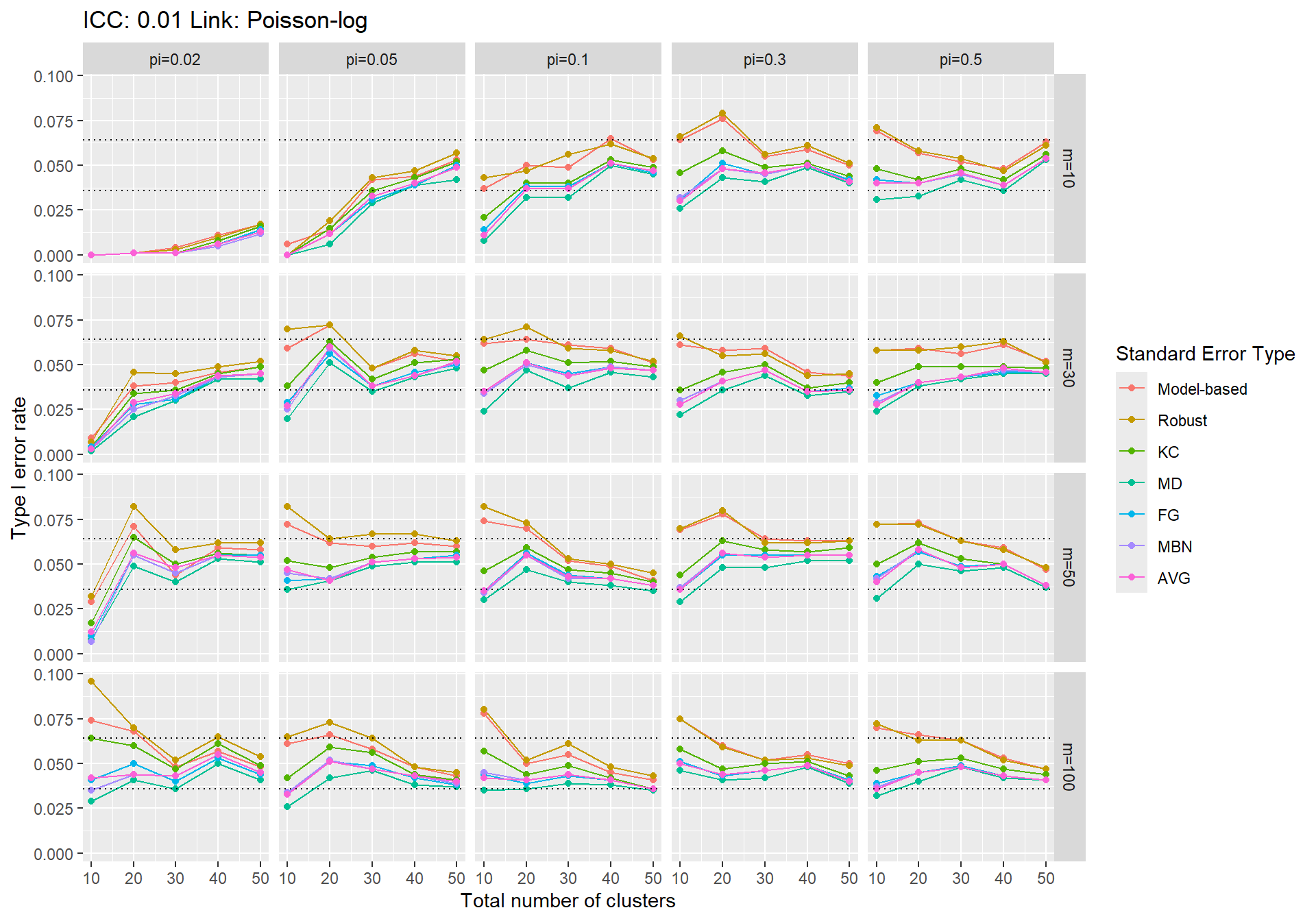}
        } 
    
    \subfloat[Poisson model with log link, ICC=0.1]{%
        \includegraphics[clip, width=0.9\textwidth]{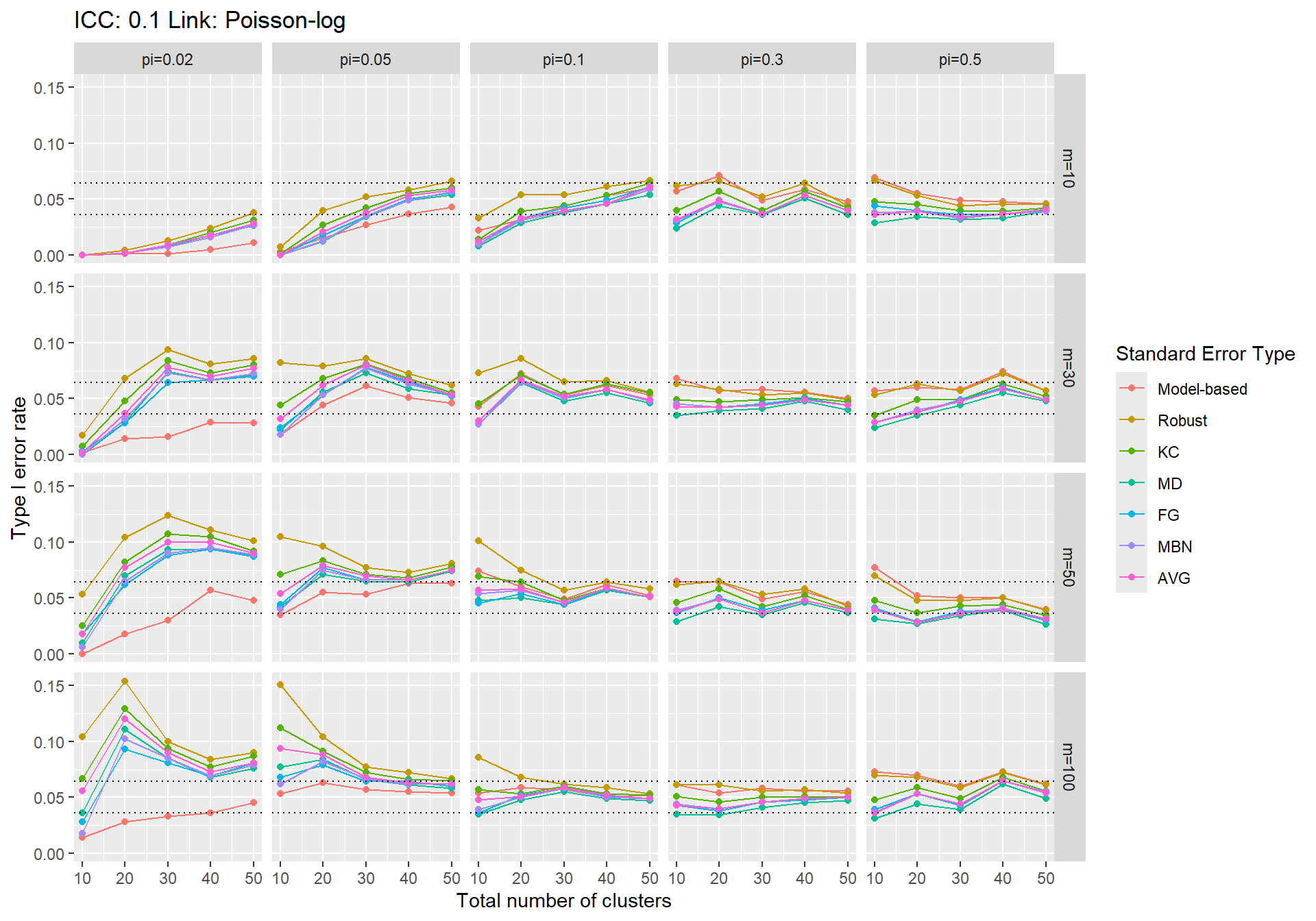}
        }
    \caption{Empirical type I error rates of GEE analyses using Poisson model to estimate risk ratios in balanced clustered data}
    \label{fig:my_label}
    \centering
\end{figure}
\begin{figure}[htp]
    \centering
    \subfloat[Gaussian model with identity link, ICC=0.01]{%
        \includegraphics[clip, width=0.9\textwidth]{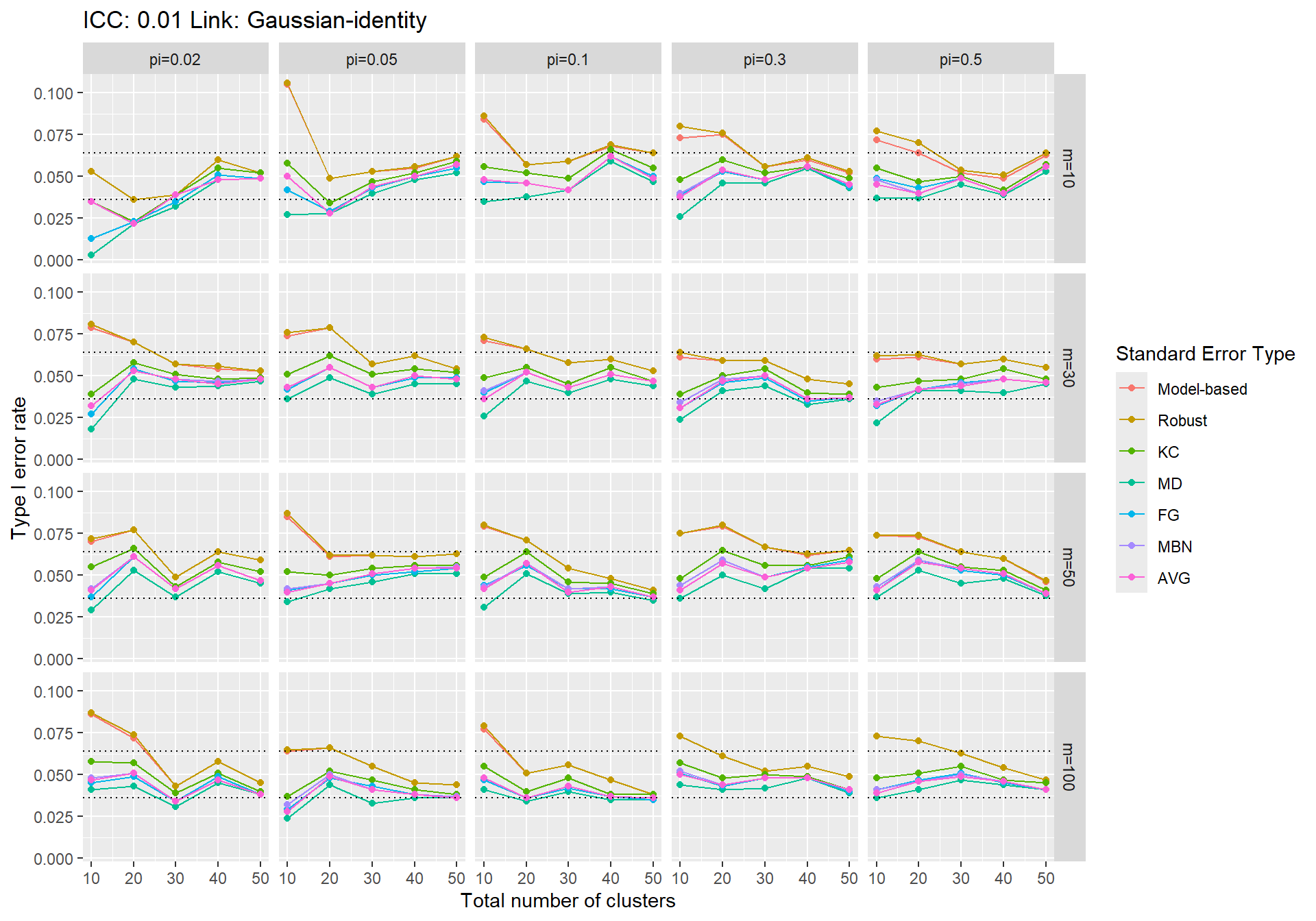}
        }
        
    \subfloat[Gaussian model with identity link, ICC=0.1]{%
        \includegraphics[clip, width=0.9\textwidth]{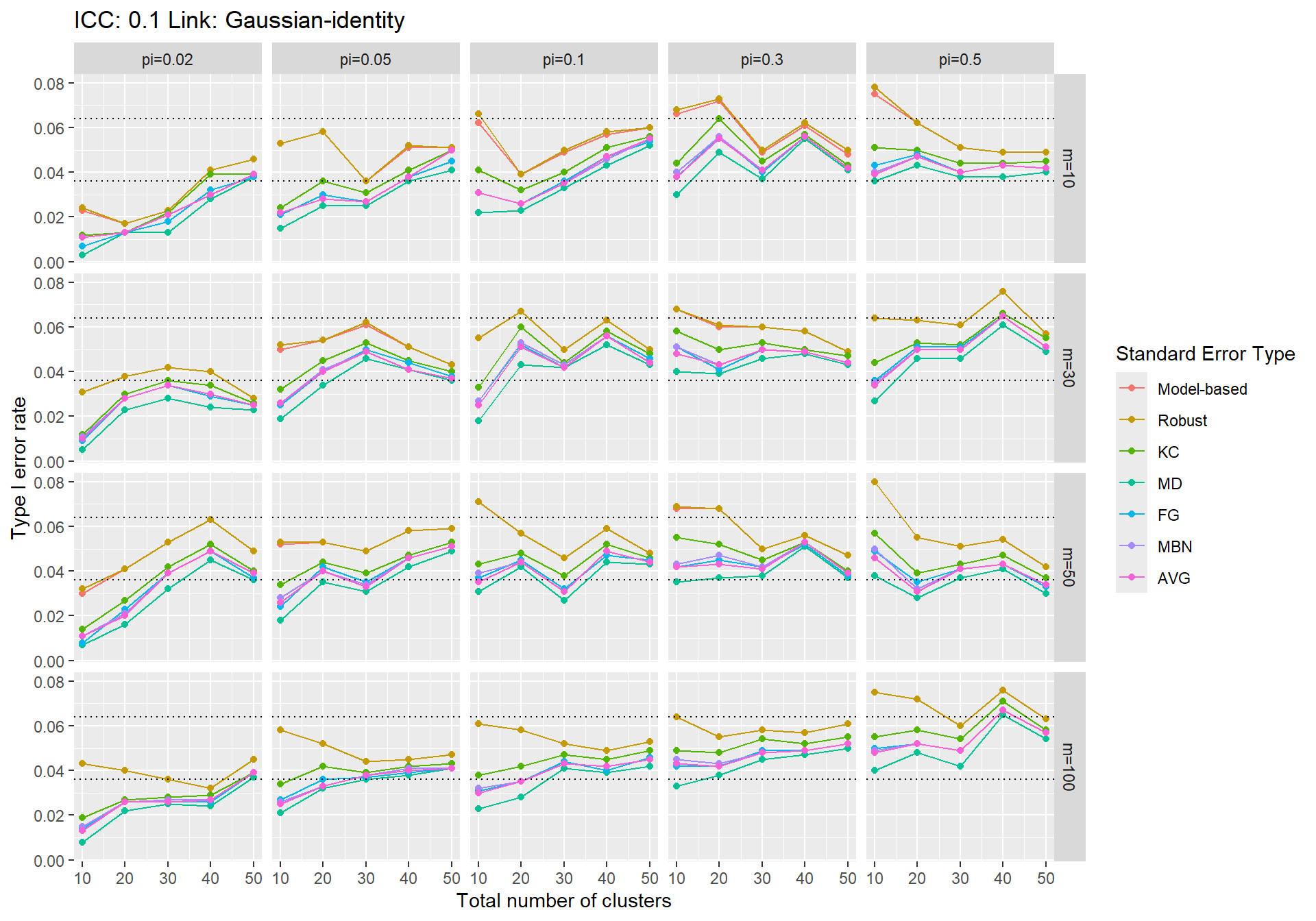}
        }
    \caption{Empirical type I error rates of GEE analyses using Poisson model to estimate risk differences in balanced clustered data}
    \label{fig:my_label}
    \centering
\end{figure}
\begin{figure}[htp]
    \centering
    \subfloat[Poisson model with log link]{%
        \includegraphics[clip, width=0.9\textwidth]{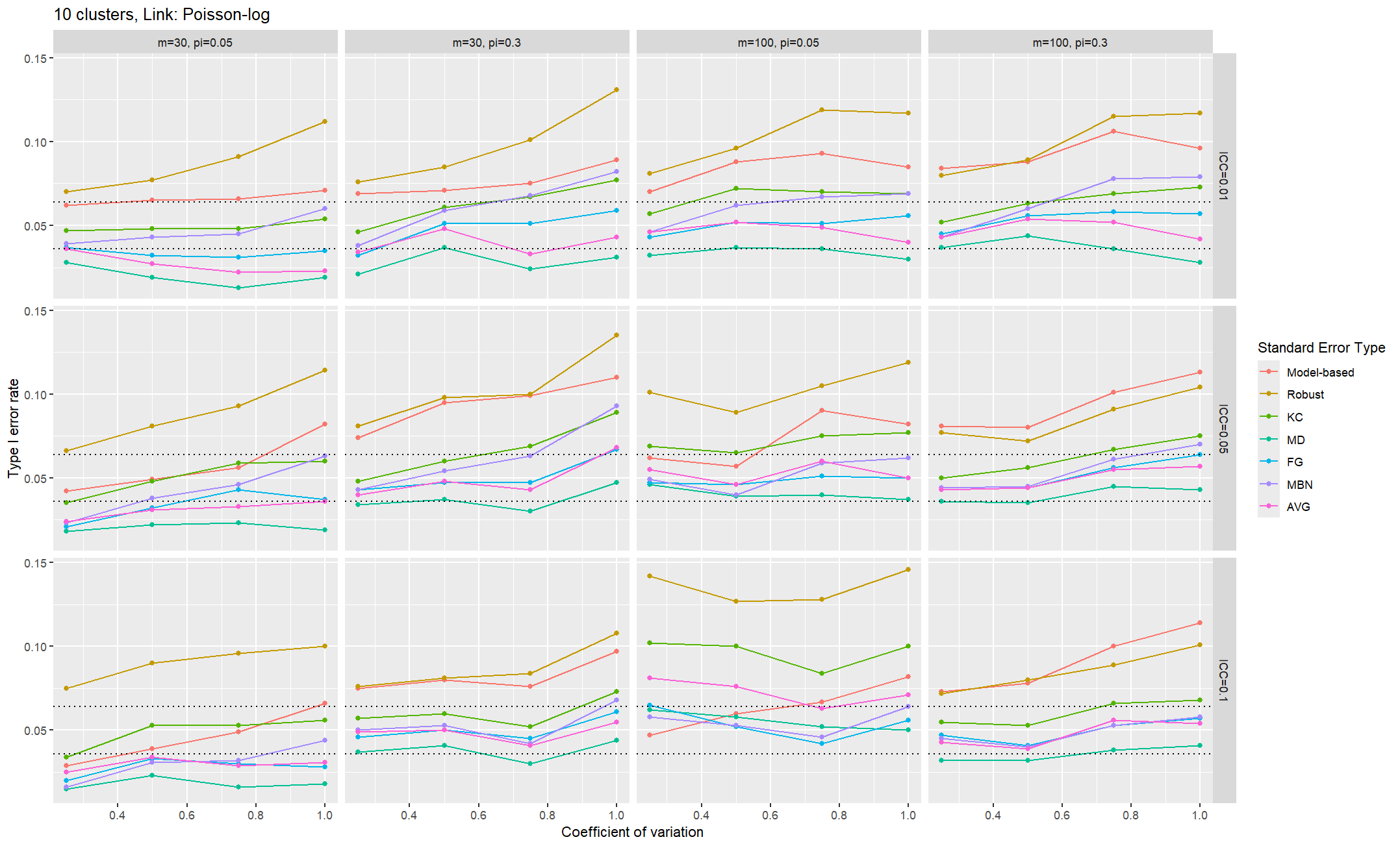}
        }

    \subfloat[Poisson model with log link]{%
        \includegraphics[clip, width=0.9\textwidth]{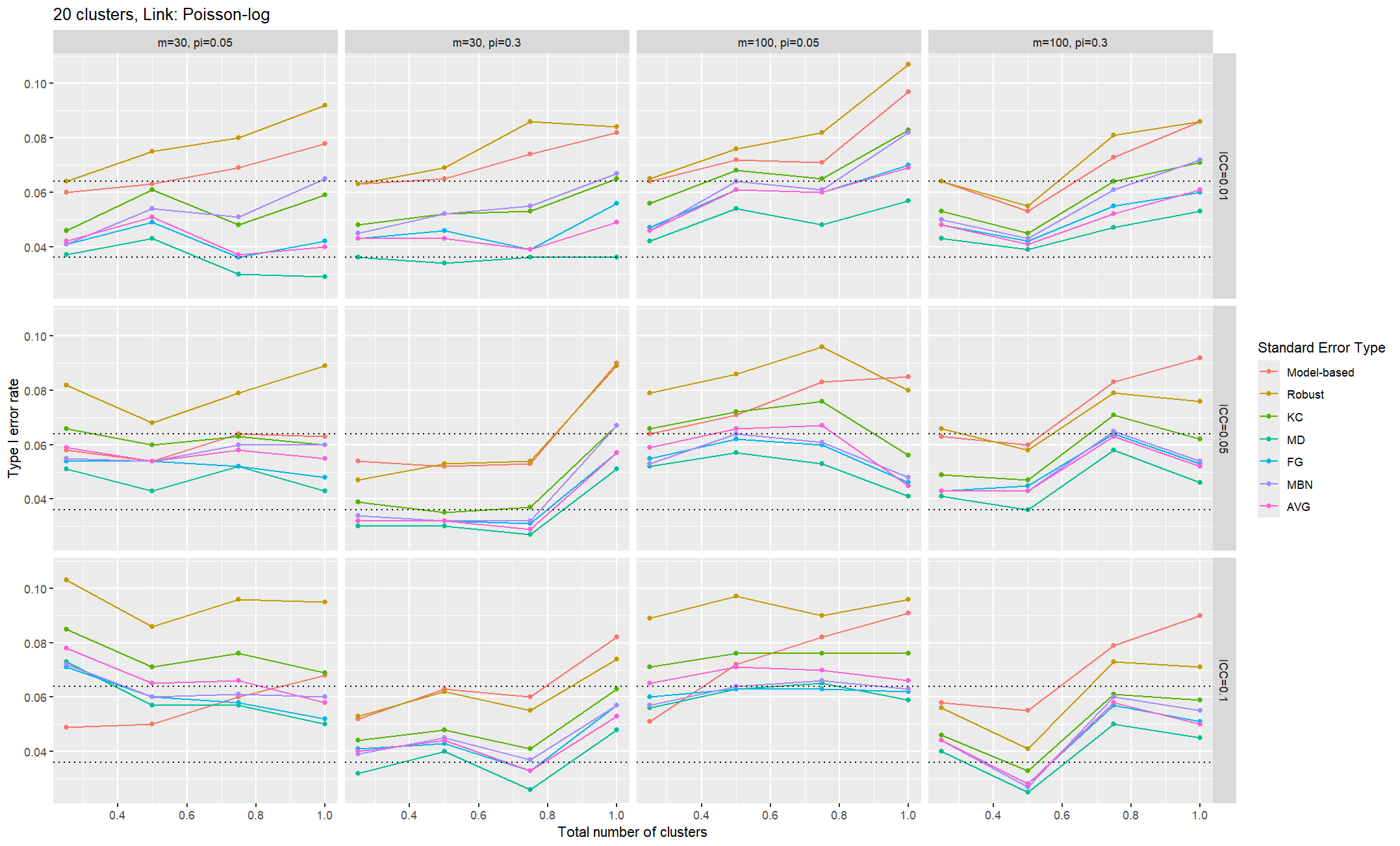}
        }
    \caption{Empirical type I error rates of GEE analyses using Poisson model to estimate risk ratios in unbalanced clustered data}
    \label{fig:my_label}
    \centering    
\end{figure}
%

\begin{figure}[htp]
    \centering
    \subfloat[Poisson model with identity link]{%
        \includegraphics[clip, width=0.9\textwidth]{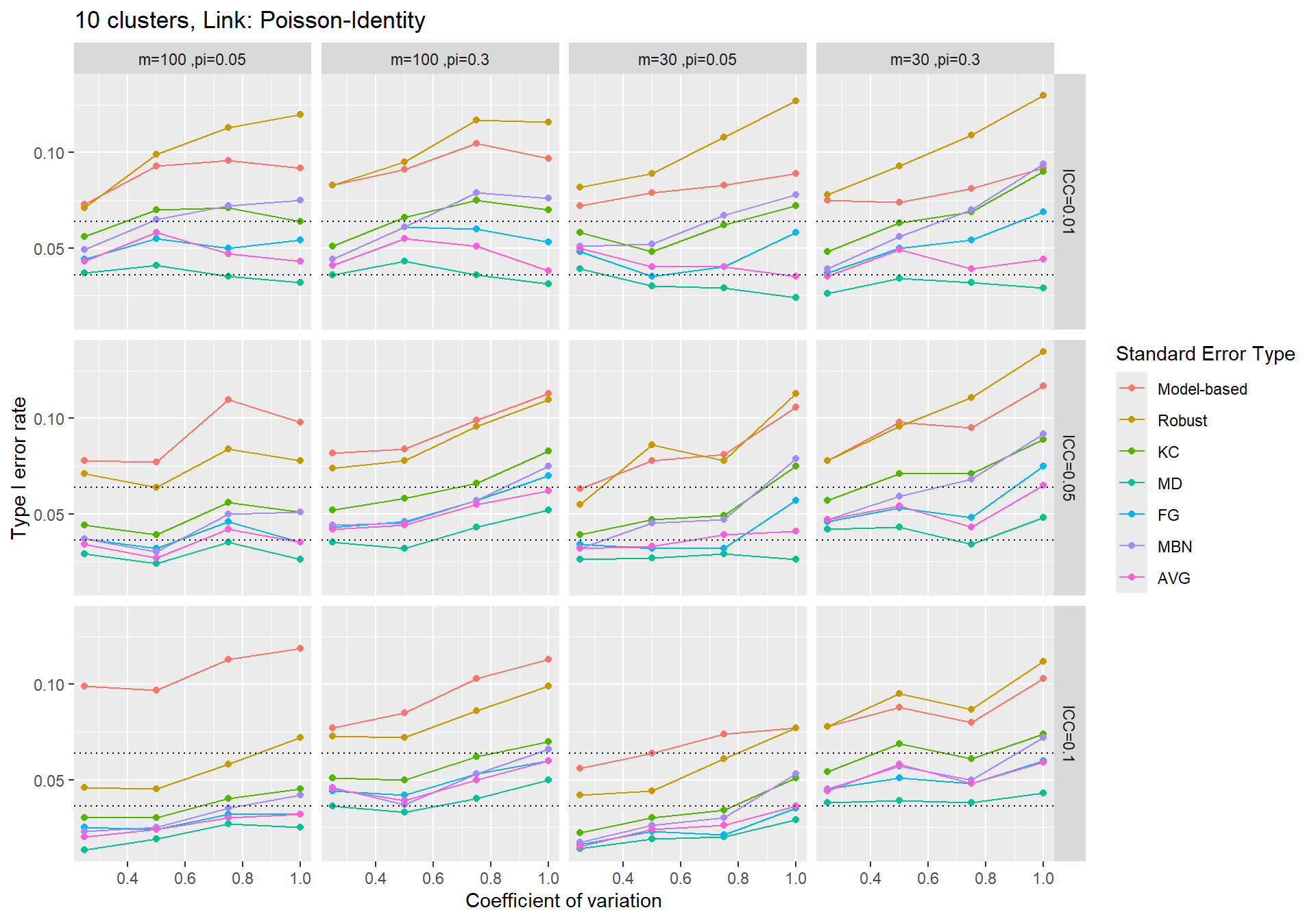}
        }

    \subfloat[Poisson model with identity link]{%
        \includegraphics[clip, width=0.9\textwidth]{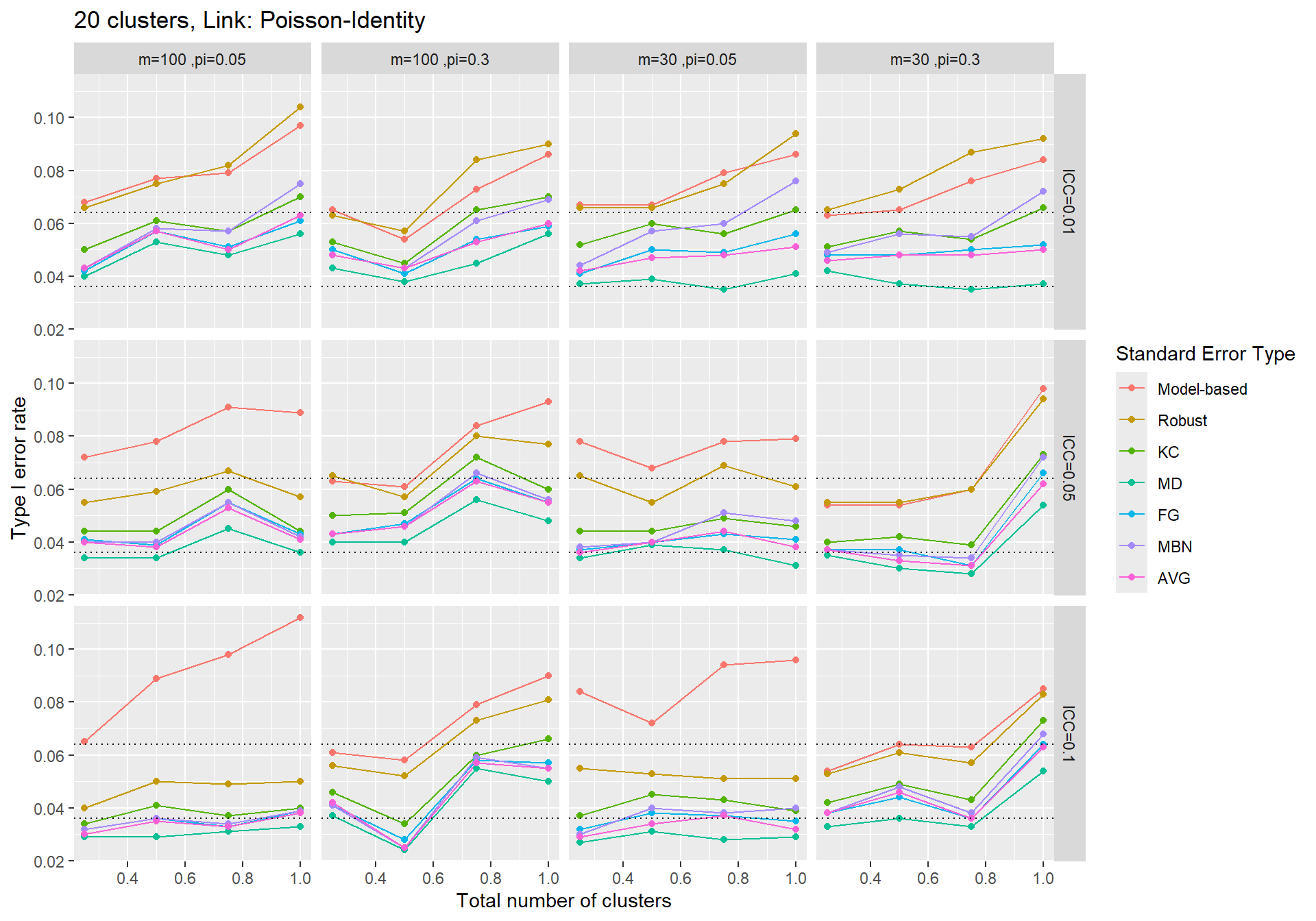}
        }
    \caption{Empirical type I error rates of GEE analyses using Poisson model to estimate risk differences in unbalanced clustered data}
    \label{fig:my_label}
    \centering    
\end{figure}

\section{Application}

In order to assess the impact of ignoring small sample correction in the analysis of real data, we drew a sample of 40 clusters from SINEMA trial. This trial aimed to evaluate the effectiveness of a community-based mobile health intervention on stroke management in China \cite{yan2021effectiveness}. The original trial enrolled 50 clusters from 5 towns with a 1:1 ratio of being randomized to the intervention arm or the control arm. The cluster size was variable across clusters with the average 24.5 participants per cluster. A binary outcome included in the trial was timed up and go (dichotomized by if the time of completion $\geq$ 14s), while some participants missed this outcome data. Based on these facts, we drew 10 clusters from 5 towns with an even proportion of intervention arm and control arm clusters, and then kept participants from each cluster with complete data. GEE Poisson models with either log link or identity link were fitted to the sampled datasets to estimate risk ratios and risk differences. The 95\% confidence intervals and p-values were provided with both the robust standard errors and KC-corrected standard errors. All the models were fitted using geeM package in Rstudio 4.1.1, and replicated using the xtgeebcv command in STATA 18.

The results are shown in Table 3. In 10 villages sampled, there were 255 participants with the complete data of the timed up and go measure, with the average number of participants per village 25.5 (CV = 0.10). The unadjusted percentage of participants with more the outcome were 30.5\% and 48.0\% for the intervention and control arms, respectively. After fitting the GEE Poisson model including the village as the cluster, the ICC was estimated to be 0.01. People receiving the intervention were estimated to have 37 percent lower risk relatively and 17.9 percent lower risk absolutely compared with the control arm. Both measures showed statistical significance (RR: P=0.035, 95\%CI [0.41, 0.96]; RD: P=0.040, 95\%CI [-0.35, -0.01]), when the robust standard errors were used. However, the corrected P-values and confidence intervals provided different evidence (RR: P=0.052, 95\%CI [0.39, 1.00]; RD: P=0.059, 95\%CI [-0.37, 0.01]). This example demonstrated that caution is required when drawing inference based on RRs and RDs in small sample settings, and appropriate correction methods should be used. 

\begin{table}[htbp]
    \begin{adjustbox}{max width=\textwidth}
    \begin{threeparttable}
    \caption{Intervention effects on stroke-related binary outcome in SINEMA}
    \label{tab:my_label}
    \centering
    \begin{tabular}{|c|c|c|c|c|c|c|c|c|}
    \hline
    \multirow{2}{*}{Outcomes} & \multicolumn{2}{c|}{Summary Statistics} & \multicolumn{3}{c|}{Risk Ratio} & \multicolumn{3}{c|}{Risk Difference} \\ \cline{2-9} & Intervention & Control & Estimate & 95\%CI & P-value & Estimate & 95\%CI & P-value \\
    \hline
    Timed up and go \tnote{a} & 39(30.05\%) & 61(48.0\%) & 0.63 & 0.41-0.96 & 0.035 & -0.18 & -0.35--0.01 & 0.040 \\ 
    \hline
    &  &  & & 0.39-1.01* \tnote{b} & 0.052* &  & -0.37-0.01* & 0.059*\\
    \hline 
    \end{tabular}
    \begin{tablenotes}
    \item [a] A test recorded in seconds during measurement and dichotomized into binary as $\geq$ 14s (indicating lower limb mobility) versus < 14s (higher limb mobility).
    \item [b] All *marked 95\% CIs and P-values use KC-corrected standard errors.
    \end{tablenotes}        
    \end{threeparttable}
    \end{adjustbox}
\end{table}

\section{Discussion}

While the odds ratio is the most widely reported effect measure in cluster randomized trials with binary outcomes, the relative risk and risk difference are also important measures and therefore appropriate methods are needed to estimate these measures and to provide valid inference in small cluster randomized trials \cite{turner2021completeness}. Given its advantage in higher convergence compared to the binomial model without loss of precision, the use of the modified Poisson approach within the GEE framework has been recommended to estimate the risk difference by using an identity link and the relative risk by using the log link  \cite{yelland2011performance} \cite{zou2004modified} \cite{pedroza2016performance} \cite{pedroza2017estimating}. When the total number of clusters is low, robust standard errors provided by GEE would be negatively biased, and a comparative evaluation of the performance of different correction methods has not been fully investigated for the GEE modified Poisson approach. In our study, the KC method had the best performance among all corrections regardless of the variation in cluster sizes, aligned with evidence from the GEE binomial approach using logit link \cite{li2015small} \cite{teerenstra2010sample}. The MD method has been previously studied alone for the analysis of the GEE modified Poisson approach \cite{yelland2011performance} \cite{zou2004modified} \cite{pedroza2016performance}, but our comparative study showed that it may reduce power due to its conservativeness. We note too that the FG method, MBN method and AVG method have similar issues. When the cluster sizes greatly vary, there is little evidence that the FG method would be superior to the KC method, which is different to the finding from other GEE models \cite{li2015small}.

Few studies considered rare outcomes in their simulation designs. In our study, we also evaluated the performance of scenarios with outcome proportion of 0.05, and found that performance of bias-corrected standard errors became less stable. While the KC method was still able to maintain the empirical type I error within the nominal range for low ICC values, it did not work as well with higher ICC. A closer association with number of clusters was also observed. Therefore, care should be taken when choosing bias correction methods in the rare outcome setting. In addition, as the odds ratio is closer to the risk ratio when the outcome risk is rare, as expected, this finding also applies to GEE logit-binomial analysis. Another factor to consider is the validity of the use of GEE model in small sample CRT with rare outcomes. Previously studies in longitudinal data showed the penalized GEE could be an alternative to GEE to improve the convergence rate and reduce the bias of regression parameters \cite{mondol2019bias} \cite{gosho2023comparison}, and thus the proper standard error estimators needs to be paired with the penalized GEE to reach valid inferences. 

Although the Poisson model is currently  not the most popular choice to estimate the risk difference, our study showed it had decent convergence rate compared to the Gaussian model when there are more than 10 clusters, which has been supported by Pedroza et al. \cite{pedroza2016performance}. The KC approach is also recommended for adjusting the robust standard error, while the latter might be even more proper for highly correlated rare outcomes. 

Given that our study was developed based on some previous findings, we did not explore every possible factor. For instance, we used the Wald $t-$test with $N-2$ degrees of freedom, as the Wald $Z-$test showed lower validity generally in the GEE binomial model with logit link \cite{li2015small}, and $N-2$ degrees of freedom is universally acceptable \cite{ford2017improved}. Our study was also limited to balanced cluster size and parallel two-arm CRT designs. It was previously shown in GEE binomial model with logit link that large variation in cluster size may affect the performance of bias correction methods \cite{li2015small}. Stepped wedge cluster trials have been increasingly used, and suitable bias correction methods to maintain the validity of inference might be different compared to the parallel-arm trial design of our study \cite{ford2020maintaining}. In addition, besides validity of inference, statistical power is also crucial for choosing the right standard error correction methods, and the advantage of Poisson model in convergence rate is more obvious when there is true treatment effect, which is the case when power was assessed rather than our assessment of Type I error performance \cite{yelland2011performance}. In addition, in spite of the nominal type I error rate, the small sample correction methods are possible to reduce the power when sparse events occur \cite{kim2020analysis}.   

In conclusion, our findings indicated that KC-corrected standard error was still proper when modified Poisson model was used to estimate RR and RD, and thus was the most robust among all estimators. The rare outcome has a noted impact on the validity of inference and requires more specific correction methods with respect to ICC and total number of clusters. Future work would be valuable in assessing performance for RR and RD estimation in CRTs with variable cluster size and in other, more complex CRT designs such as SW-CRTs.

\section*{Acknowledgements} 
The authors wish to thank members of the Research Design and Analysis Core of the Duke Global Health Institute for feedback on this work and for permission from Professor Lijing Yan for permission to use the Sinema data. Partial funding for this study was provided by three awards from the National Institutes of Health: the RESHAPE trial (R01MH120649) and the Bachpan study (R01HD075875) and the TESTsmART trial (R01DK109696).

\section*{Supplementary Material}
Supplementary material include additional figures summarizing results of the simulation study. Data and code is available from the first author.

\clearpage
\newpage
\bibliographystyle{unsrtnat}
\bibliography{references}  






\newpage
\appendix

\section{Supplemental Materials}
In this section, we present additional simulation results for balanced clustered data under a range of models and effect measures. Supplementary Figures 1–2 display the relative bias and Type I error of GEE analyses using a Poisson model to estimate risk ratios when the ICC is 0.05, and Supplementary Figures 3–4 provide corresponding results for risk differences. Supplementary Figures 5–7 and 8–10 summarize the Type I error of GEE analyses using a Binomial model to estimate risk ratios and risk differences, respectively. Supplementary Figures 11–13 report the Type I error of GEE analyses using a Gaussian model to estimate risk differences, while Supplementary Figures 14–16 present the Type I error of GEE analyses using a Binomial model to estimate odds ratios.

\setcounter{figure}{0}
\renewcommand{\thefigure}{S\arabic{figure}}   
\renewcommand{\figurename}{Supp Figure}       

\begin{figure}[htp]
    \centering
    \subfloat[Poisson model with log link, ICC=0.05]{%
        \includegraphics[clip, width=0.48\textwidth]{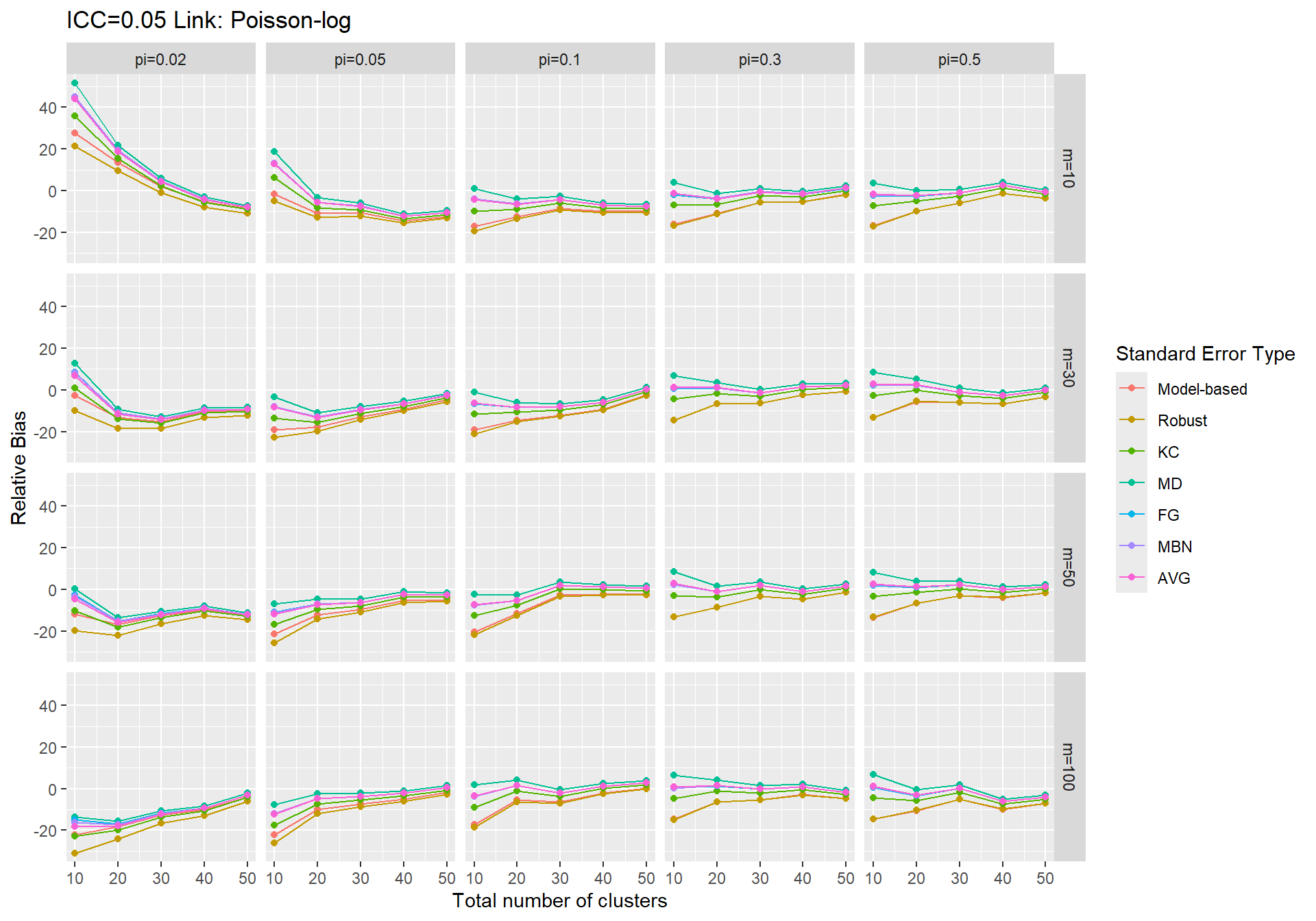}
        } 
                 \hfill
    \subfloat[Poisson model with log link, ICC=0.05]{%
        \includegraphics[clip, width=0.48\textwidth]{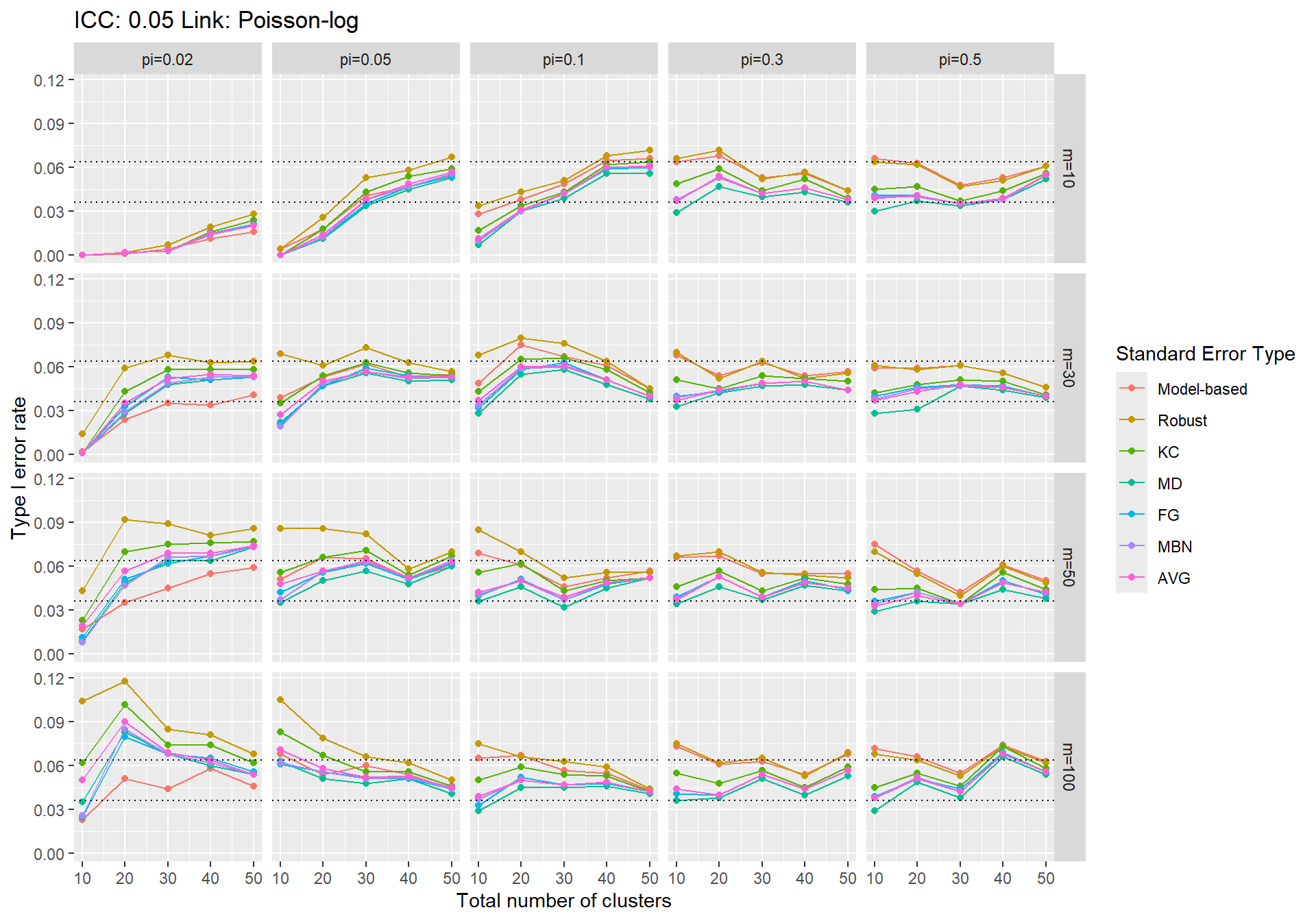}
        } 
    \caption{Relative Bias and Type I Error of GEE analyses using Poisson model to estimate risk ratios in balanced clustered data under ICC of 0.05 }
    \label{fig:my_label}
    \centering
\end{figure}

\begin{figure}[htp]
    \centering
    \subfloat[Poisson model with identity link, ICC=0.01]{%
        \includegraphics[clip, width=0.48\textwidth]{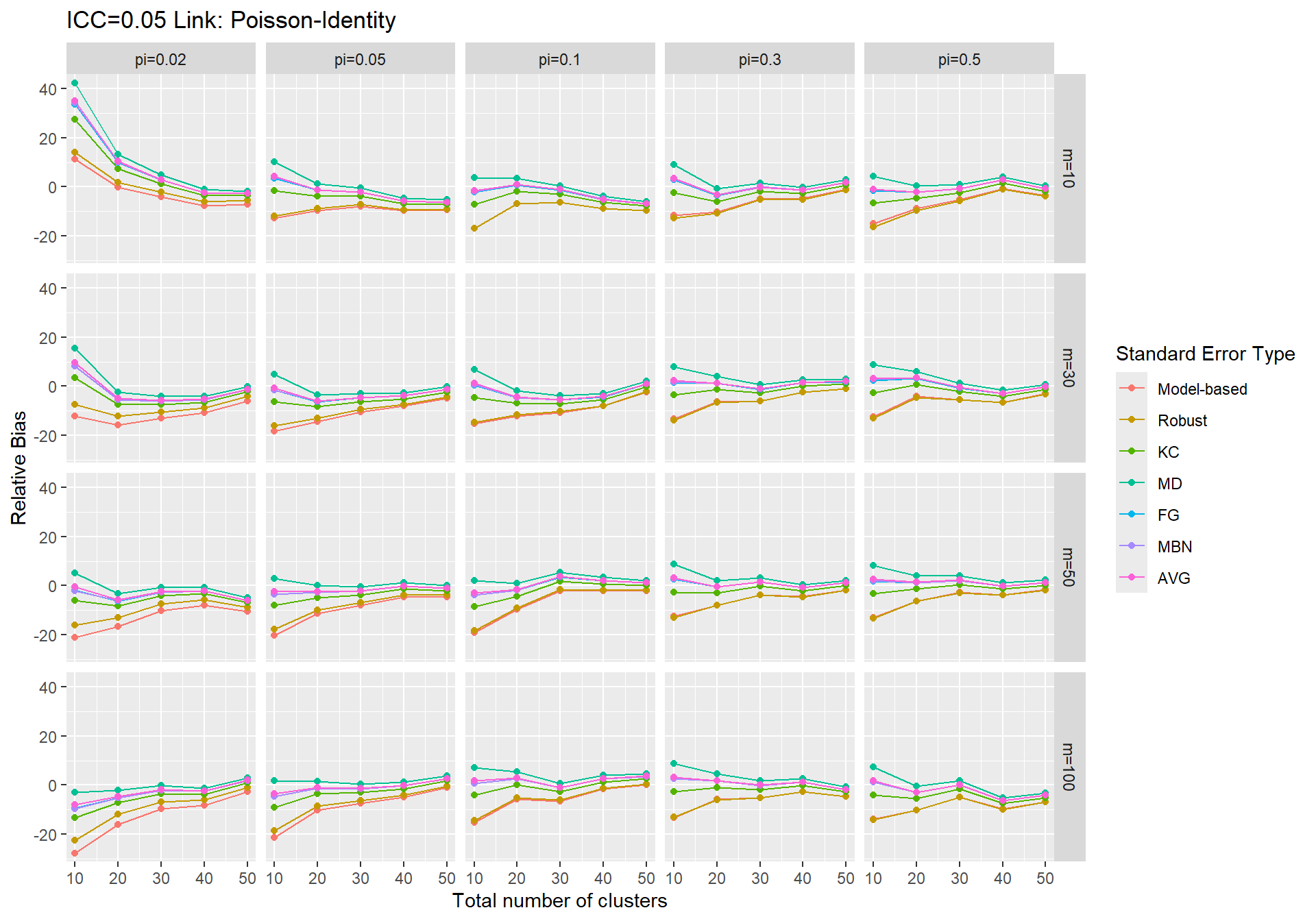}
        } 
                 \hfill
    \subfloat[Poisson model with log link, ICC=0.05]{%
        \includegraphics[clip, width=0.48\textwidth]{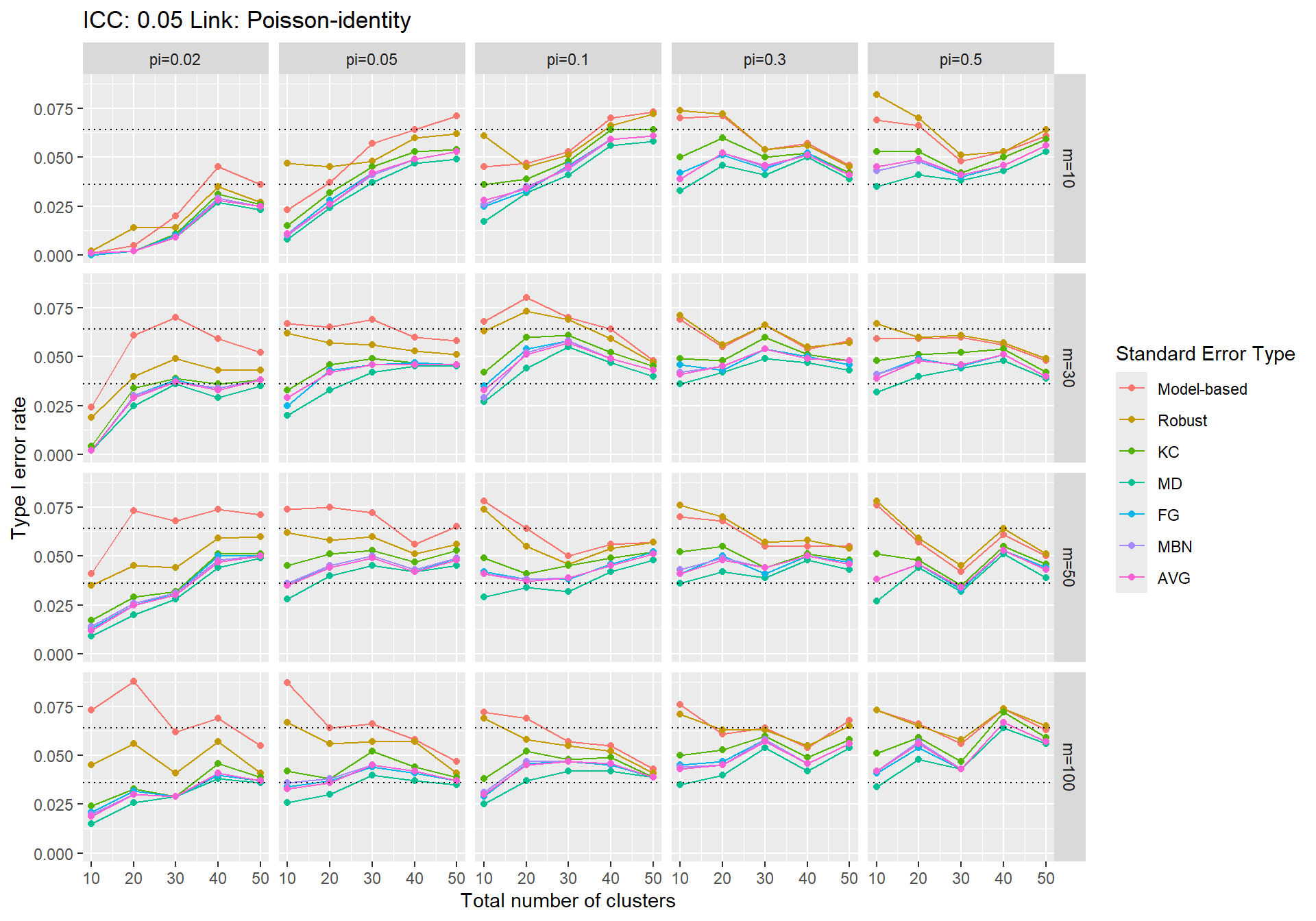}
        } 
                 \hfill
    \caption{Relative Bias and Type I Error of GEE analyses using Poisson model to estimate risk differences in balanced clustered data under ICC of 0.05 }
    \label{fig:my_label}
    \centering
\end{figure}

\begin{figure}[htp]
    \centering
    \subfloat[Binomial model with log link, ICC=0.01]{%
        \includegraphics[clip, width=0.48\textwidth]{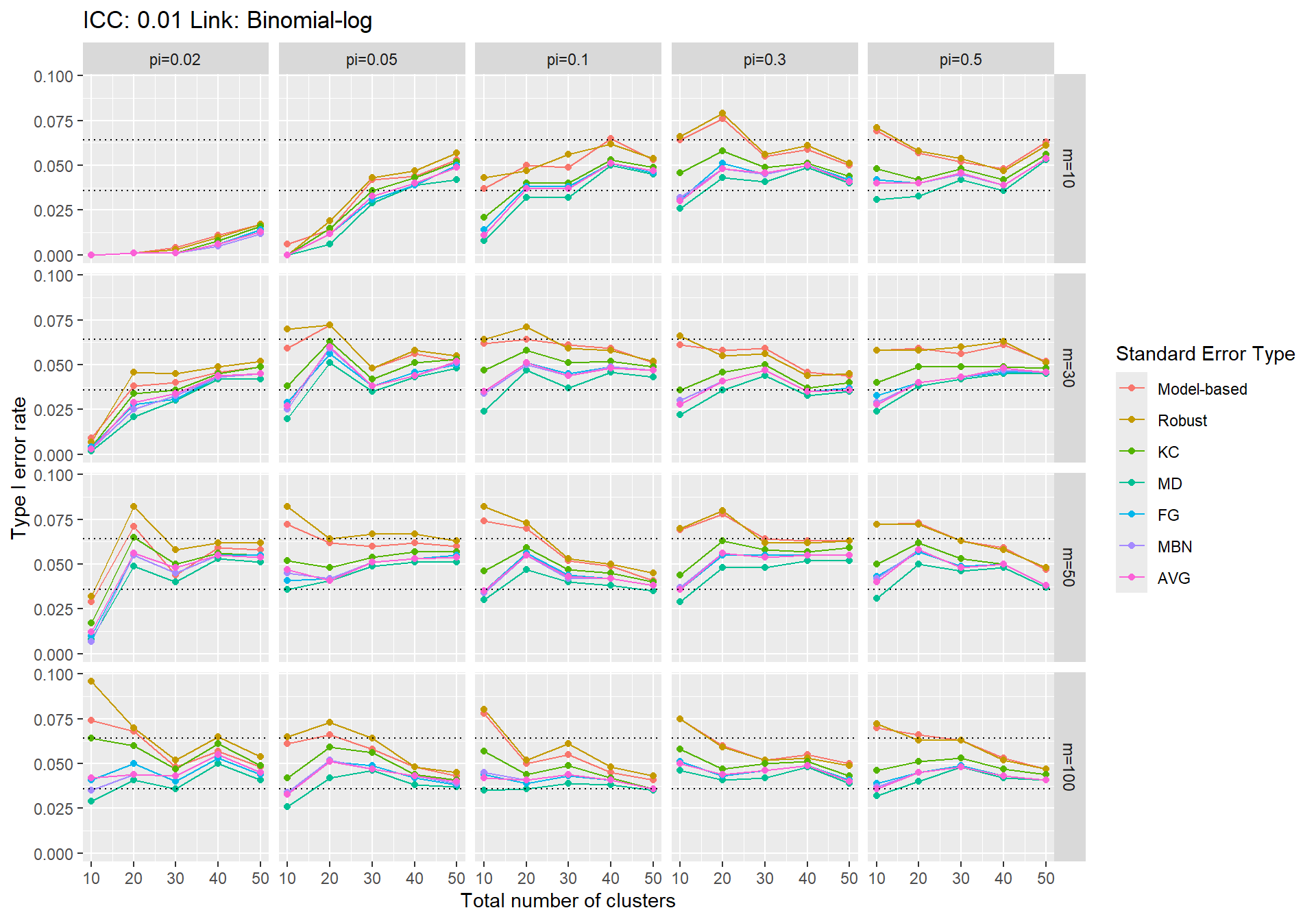}
        } 
                 \hfill
    \subfloat[Binomial model with log link, ICC=0.05]{%
        \includegraphics[clip, width=0.48\textwidth]{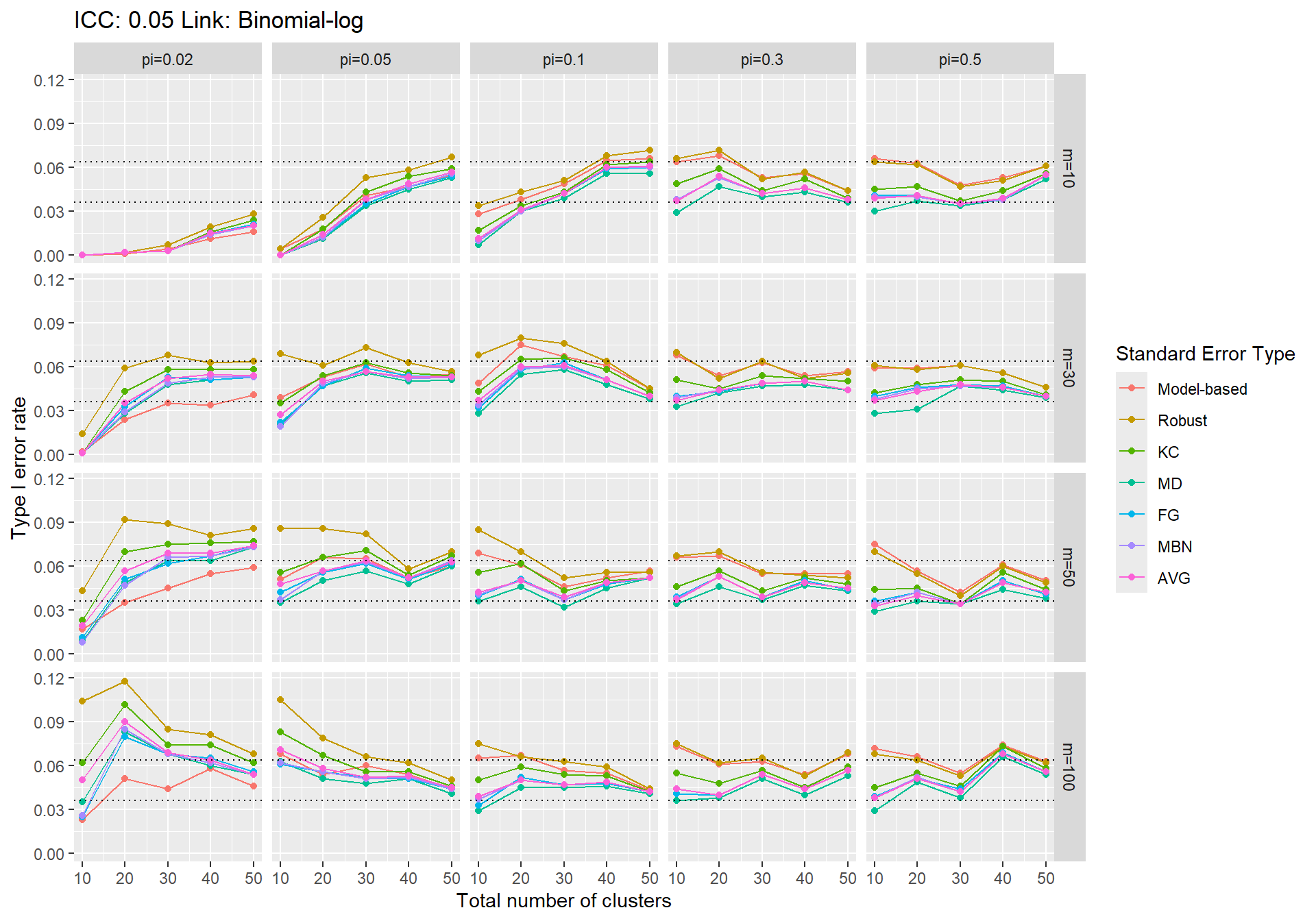}
        } 
                 \hfill
    \subfloat[Binomial model with log link, ICC=0.1]{%
        \includegraphics[clip, width=0.48\textwidth]{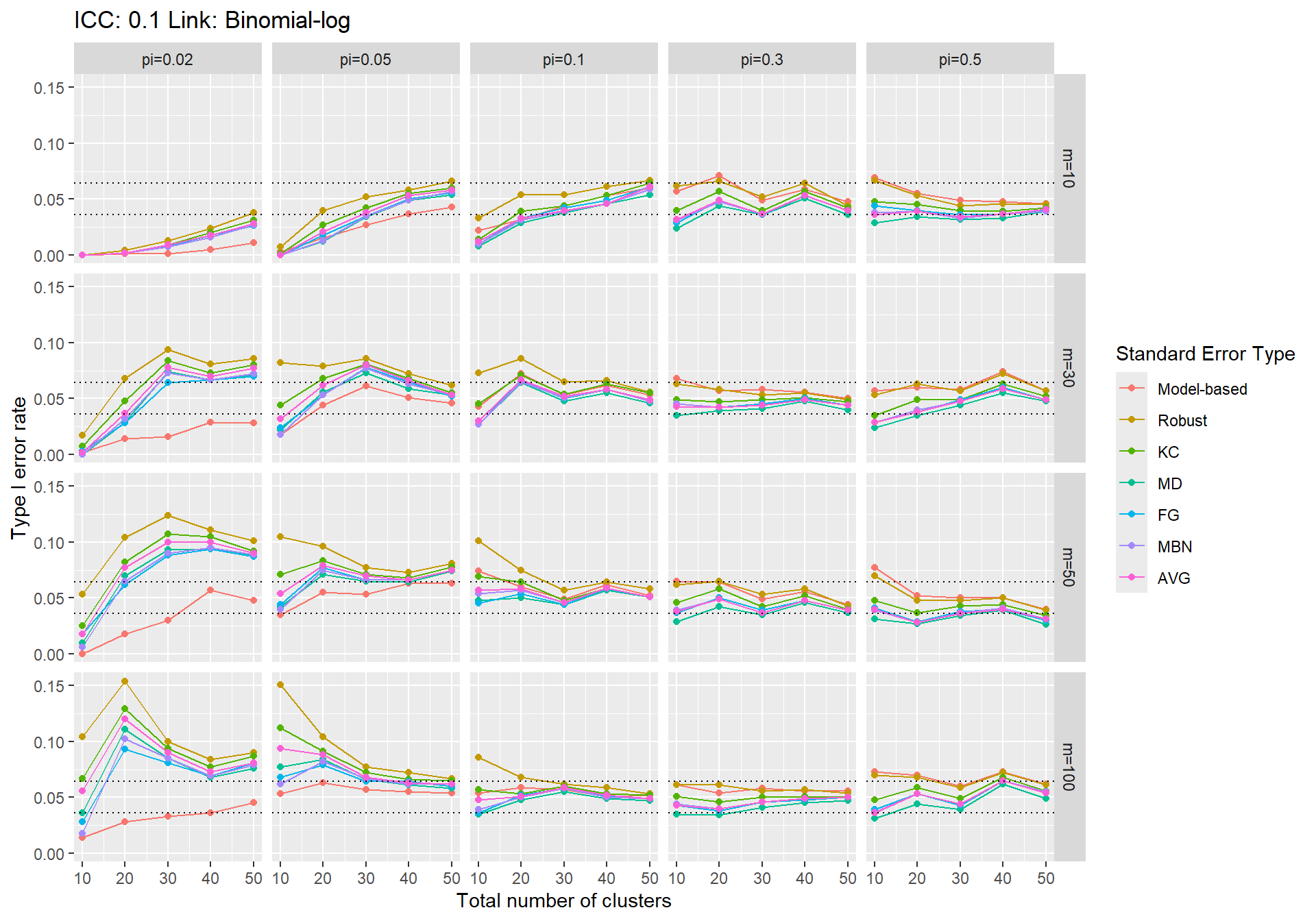}
        } 
    \caption{Type I Error of GEE analyses using Binomial model to estimate risk ratios in balanced clustered data }
    \label{fig:my_label}
    \centering
\end{figure}

\begin{figure}[htp]
    \centering
    \subfloat[Binomial model with identity link, ICC=0.01]{%
        \includegraphics[clip, width=0.48\textwidth]{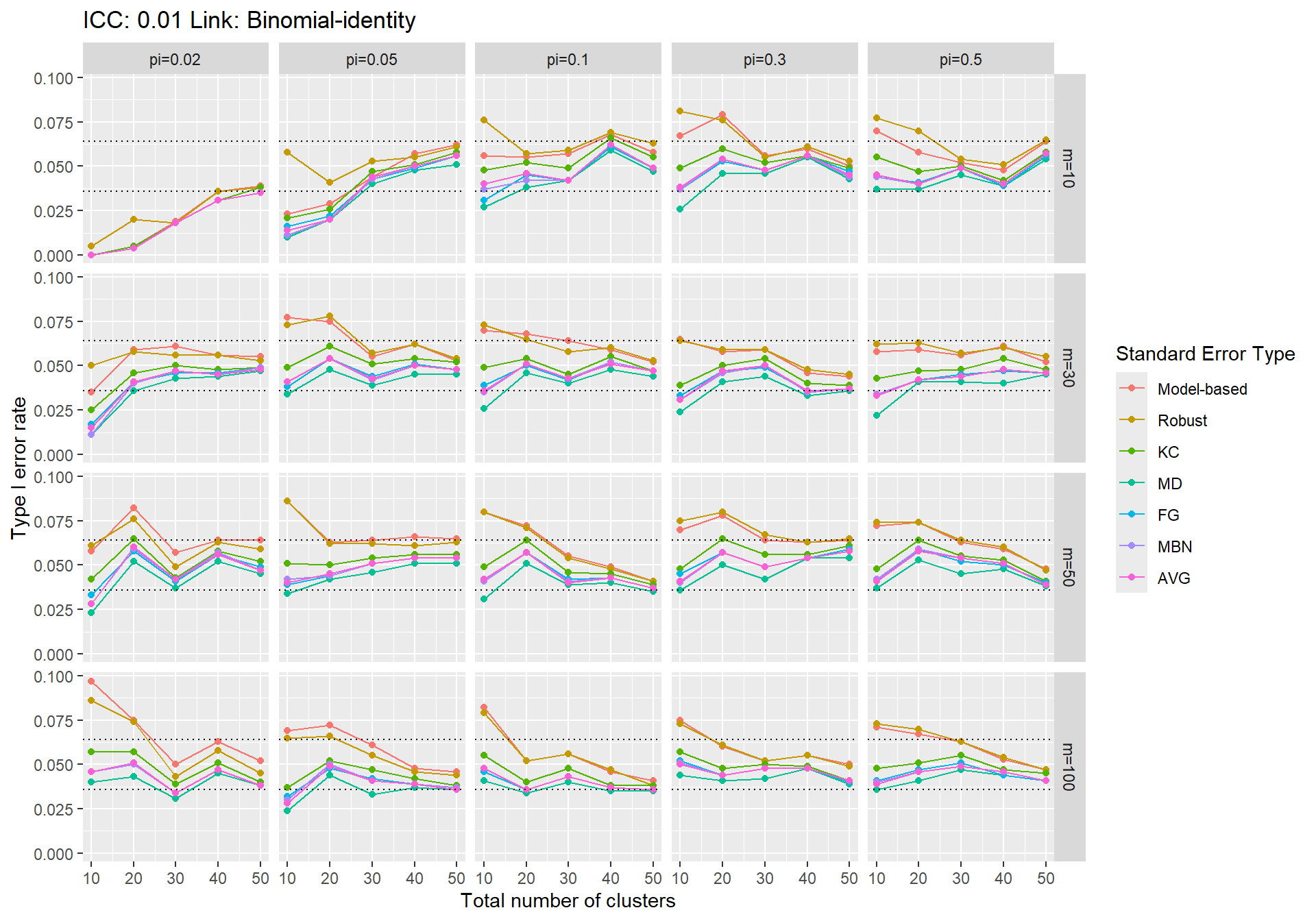}
        } 
                 \hfill
    \subfloat[Binomial model with identity link, ICC=0.05]{%
        \includegraphics[clip, width=0.48\textwidth]{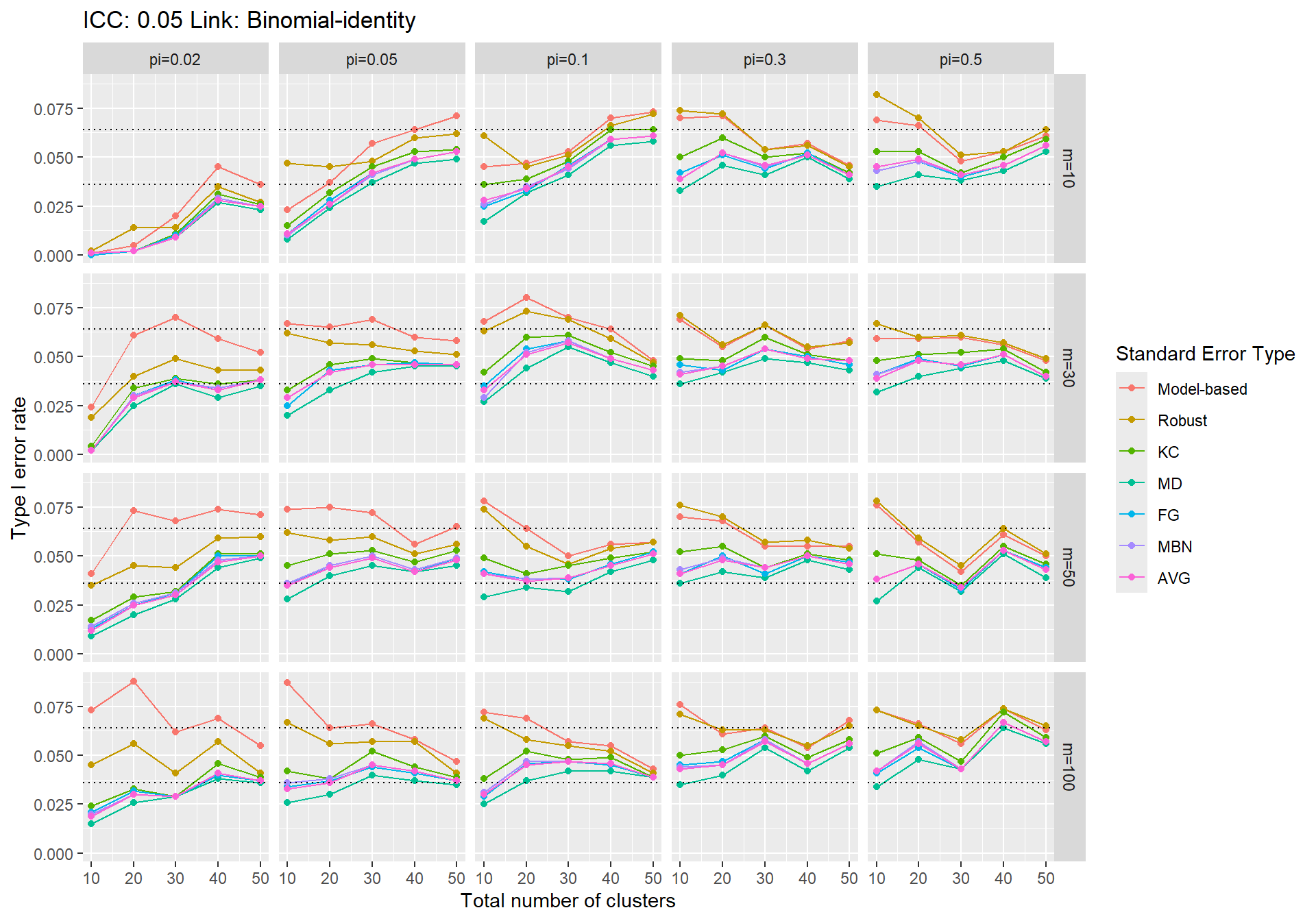}
        } 
                 \hfill
    \subfloat[Binomial model with identity link, ICC=0.1]{%
        \includegraphics[clip, width=0.48\textwidth]{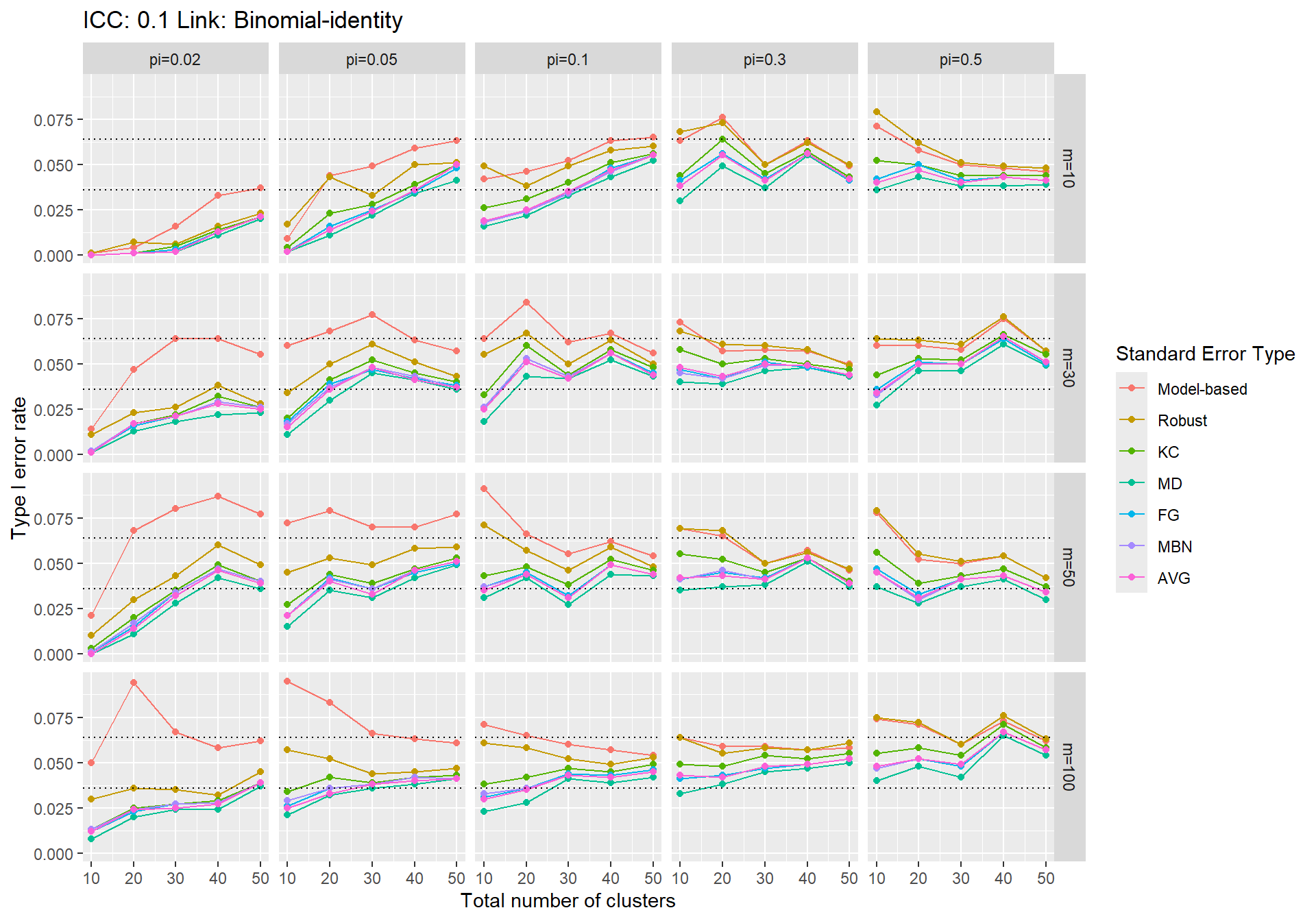}
        } 
    \caption{Type I Error of GEE analyses using Binomial model to estimate risk differences in balanced clustered data }
    \label{fig:my_label}
    \centering
\end{figure}

\begin{figure}[htp]
    \centering
    \subfloat[Gaussian model with identity link, ICC=0.01]{%
        \includegraphics[clip, width=0.48\textwidth]{figures/Image/TypeIError_GauIden_ICC001.png}
        } 
         \hfill
    \subfloat[Gaussian model with identity link, ICC=0.05]{%
        \includegraphics[clip, width=0.48\textwidth]{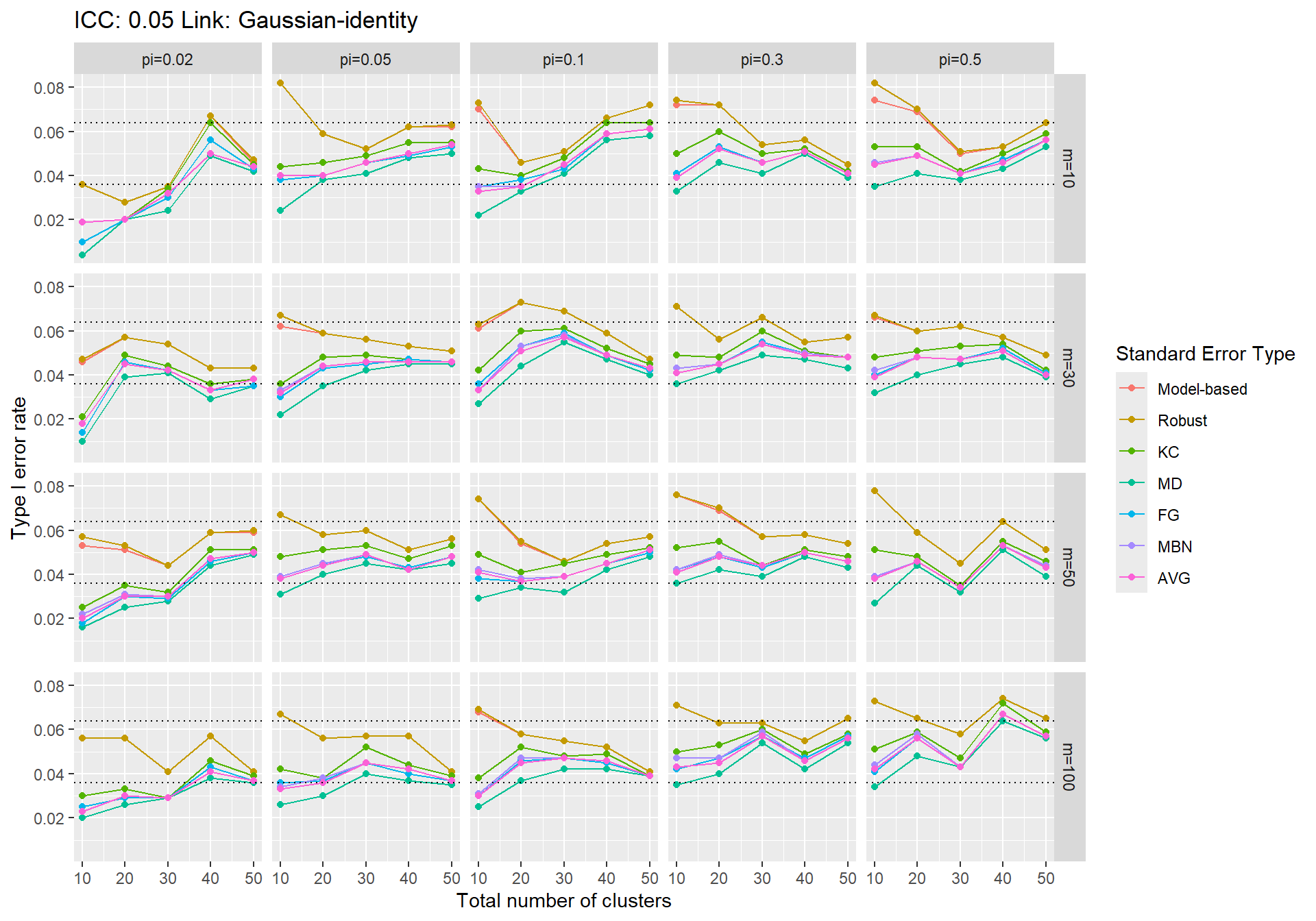}
        } 
         \hfill
    \subfloat[Gaussian model with identity link, ICC=0.1]{%
        \includegraphics[clip, width=0.48\textwidth]{figures/Image/TypeIError_GauIden_ICC010.png}
        } 
    \caption{Type I Error of GEE analyses using Gaussian model to estimate risk differences in balanced clustered data }
    \label{fig:my_label}
    \centering
\end{figure}

\begin{figure}[htp]
    \centering
    \subfloat[Binomial model with logit link, ICC=0.01]{%
        \includegraphics[clip, width=0.48\textwidth]{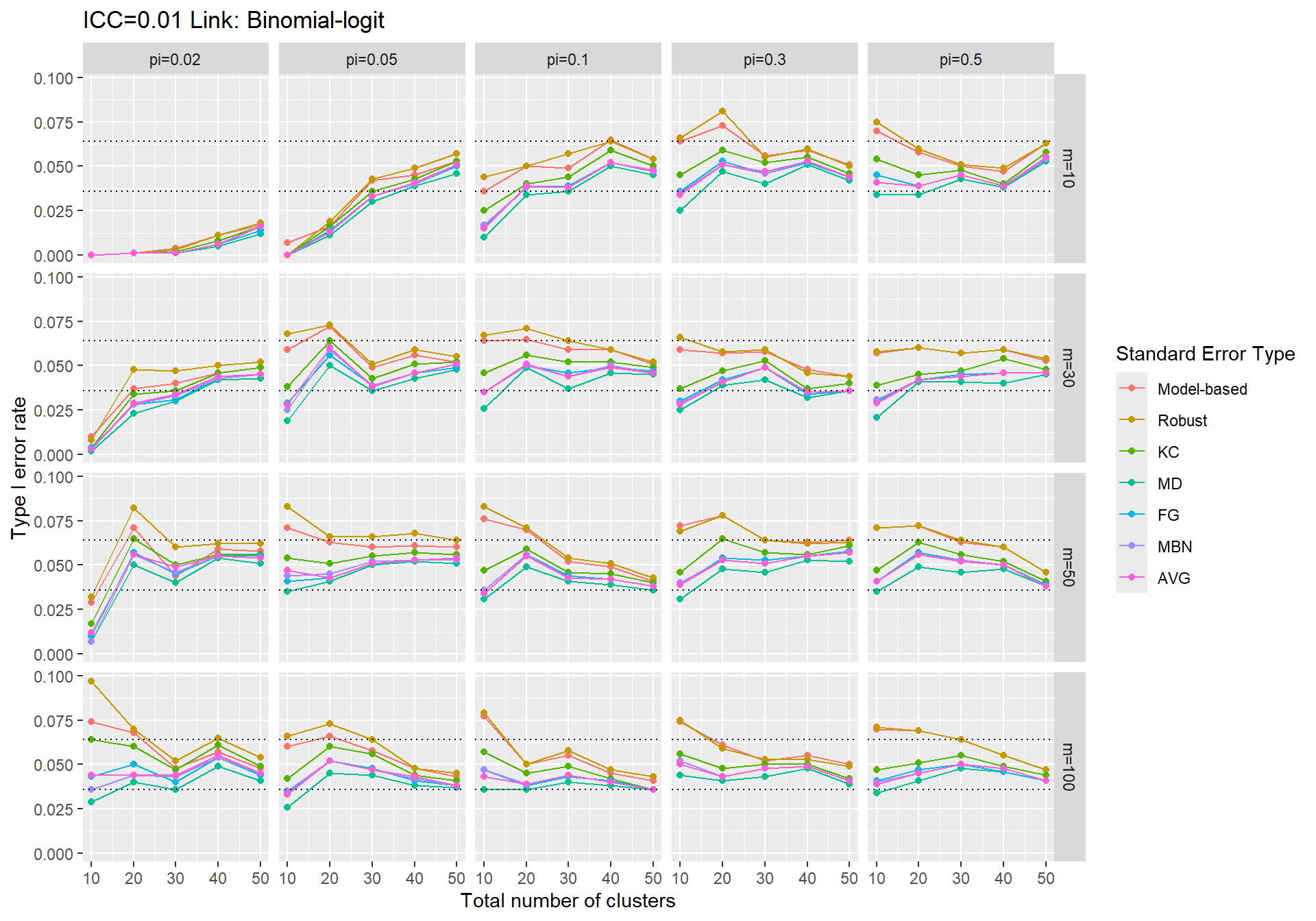}
        } 
            \hfill
    \subfloat[Binomial model with logit link, ICC=0.05]{%
        \includegraphics[clip, width=0.48\textwidth]{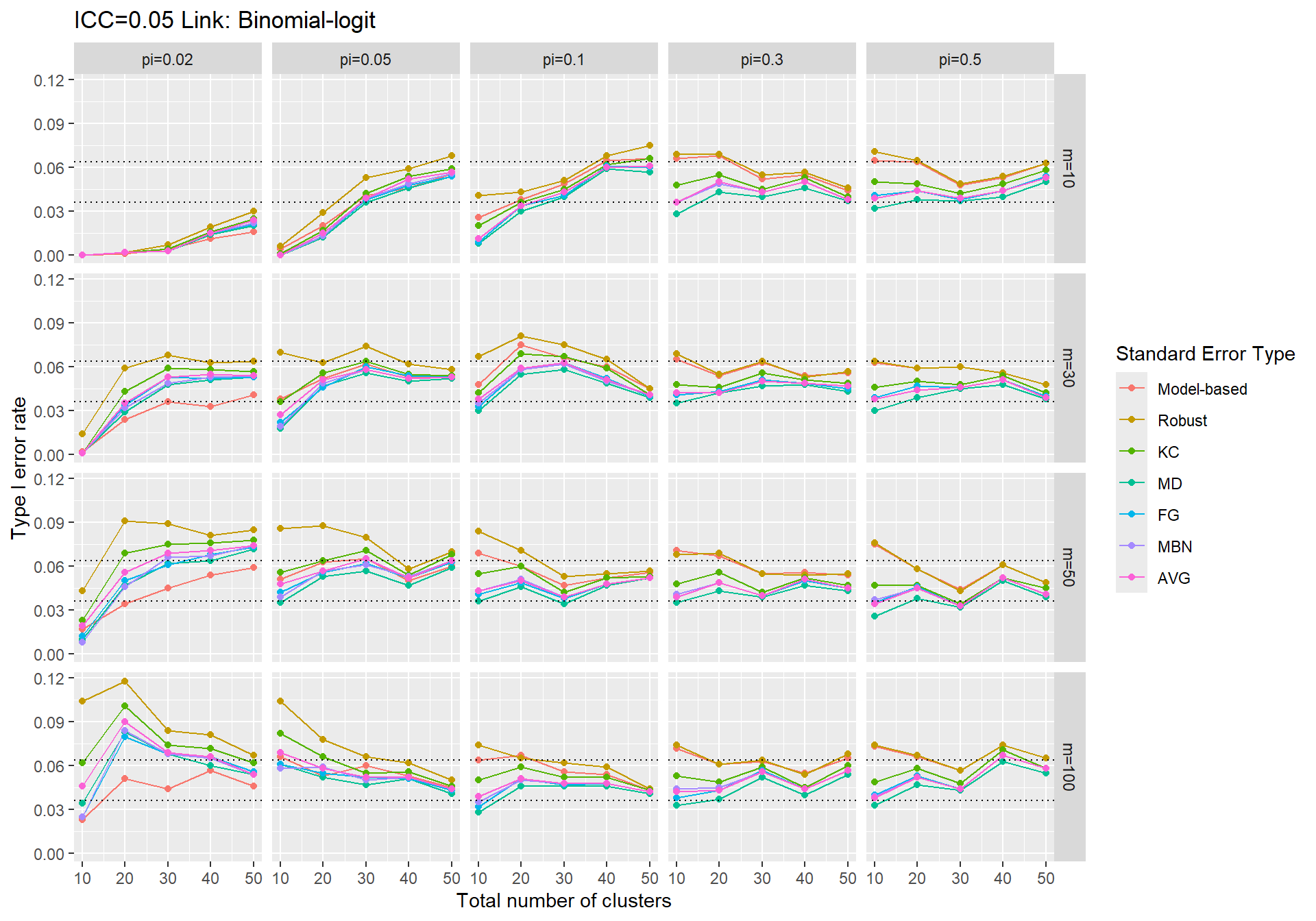}
        } 
            \hfill
    \subfloat[Binomial model with logit link, ICC=0.1]{%
        \includegraphics[clip, width=0.48\textwidth]{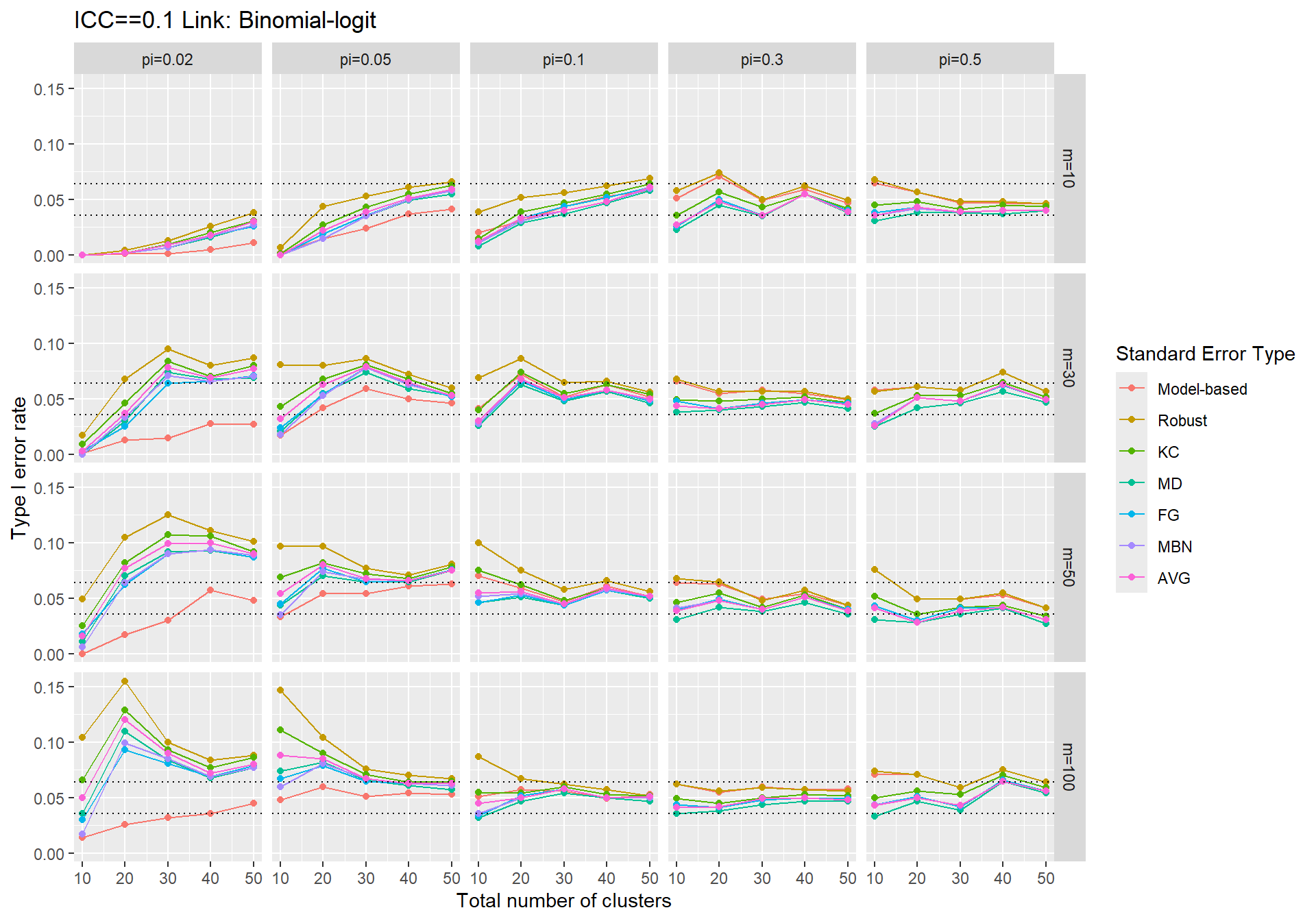}
        } 
    \caption{Type I Error of GEE analyses using Binomial model to estimate odds ratios in balanced clustered data }
    \label{fig:my_label}
    \centering
\end{figure}

\end{document}